\documentclass{aastex701}
\usepackage{multirow}
\usepackage{xcolor}
\usepackage[normalem]{ulem}

\begin{document}

\title{Advancing Electroformed X-ray Optics Bridging Synchrotron Radiation Technology and Space Astronomy: Improved Figure Accuracy and X-ray Imaging Performance}

\author[orcid=0000-0002-9901-233X]{Ikuyuki Mitsuishi} 
\affiliation{Graduate School of Science, Tokai National Higher Education and Research System, Nagoya University, Furo-cho, Chikusa-ku, Nagoya, Aichi 464-8602, Japan}
\email[show]{mitsuisi@u.phys.nagoya-u.ac.jp}  

\author[orcid=0009-0005-2776-7775]{Yusuke Yoshida} 
\affiliation{Graduate School of Science, Tokai National Higher Education and Research System, Nagoya University, Furo-cho, Chikusa-ku, Nagoya, Aichi 464-8602, Japan}
\email{yoshida_y@u.phys.nagoya-u.ac.jp}  

\author[orcid=0009-0005-3295-7215]{Kazuki Ampuku} 
\affiliation{Graduate School of Science, Tokai National Higher Education and Research System, Nagoya University, Furo-cho, Chikusa-ku, Nagoya, Aichi 464-8602, Japan}
\email{ampuku_k@u.phys.nagoya-u.ac.jp}  


\author[orcid=0009-0007-1256-4611]{Makoto Yoshihara} 
\affiliation{Graduate School of Science, Tokai National Higher Education and Research System, Nagoya University, Furo-cho, Chikusa-ku, Nagoya, Aichi 464-8602, Japan}
\email{yoshihara_m@u.phys.nagoya-u.ac.jp}

\author{Keitoku Yoshihira} 
\affiliation{Graduate School of Science, Tokai National Higher Education and Research System, Nagoya University, Furo-cho, Chikusa-ku, Nagoya, Aichi 464-8602, Japan}
\email{yoshihira_k@u.phys.nagoya-u.ac.jp}  

\author{Tetsuo Kano}
\affiliation{Graduate School of Science, Tokai National Higher Education and Research System, Nagoya University, Furo-cho, Chikusa-ku, Nagoya, Aichi 464-8602, Japan}
\email{kano.tetsuo.s6@f.mail.nagoya-u.ac.jp}  

\author{Naoki Ishida} 
\affiliation{Graduate School of Science, Tokai National Higher Education and Research System, Nagoya University, Furo-cho, Chikusa-ku, Nagoya, Aichi 464-8602, Japan}
\email{ishida.naoki.i2@f.mail.nagoya-u.ac.jp}  

\author[orcid=0009-0005-2657-2554]{Koki Sakuta} 
\affiliation{Astronomical Institute, Tohoku University, 6-3 Aramakiazaaoba, Aoba-ku, Sendai, Miyagi 980-8578, Japan}
\email{koki.sakuta.a1@tohoku.ac.jp}  

\author[orcid=0000-0002-6330-3944]{Noriyuki Narukage} 
\affiliation{National Astronomical Observatory Japan, 2-21-1 Osawa, Mitaka, Tokyo 181-8588, Japan}
\email{noriyuki.narukage@nao.ac.jp}  

\author[orcid=0000-0000-0000-0000]{Keisuke Tamura} 
\affiliation{NASA/GSFC, 8800 Greenbelt Rd, Greenbelt, MD 20771, US}
\affiliation{Center for Space Sciences and Technology, University of Maryland, Baltimore County (UMBC), Baltimore, MD, 21250 USA}
\email{ktamura1@umbc.edu}  

\author[orcid=0000-0000-0000-0000]{Kikuko Miyata} 
\affiliation{Meijo University, 1-501 Shiogamaguchi, Tempaku-ku, Nagoya, Aichi 468-8502, Japan}
\email{kmiyata@meijo-u.ac.jp}  

\author[orcid=0000-0003-4305-807X]{Gota Yamaguchi} 
\affiliation{RIKEN/ SPring-8, 1-1-1, Kouto, Sayo-cho, Sayo-gun, Hyogo 679-5198, Japan}
\email{g-yamaguchi@spring8.or.jp}  

\author[orcid=0000-0002-6841-0116]{Hidekazu Takano} 
\affiliation{RIKEN/ SPring-8, 1-1-1, Kouto, Sayo-cho, Sayo-gun, Hyogo 679-5198, Japan}
\email{h.takano@spring8.or.jp}  

\author[orcid=0000-0001-5182-063X]{Yoshiki Kohmura} 
\affiliation{RIKEN/ SPring-8, 1-1-1, Kouto, Sayo-cho, Sayo-gun, Hyogo 679-5198, Japan}
\email{yoshiki.kohmura@riken.jp}  

\author{Shutaro Mohri} 
\affiliation{School of Engineering, The University of Tokyo, 7-3-1 Hongo, Bunkyo, Tokyo 113-8656, Japan}
\email{mohri-shutaro644@g.ecc.u-tokyo.ac.jp}  

\author[orcid=0000-0002-1254-4297]{Takehiro Kume} 
\affiliation{Natsume Optical Corporation, 3461 Kamichyaya, Kanae, Iida, Nagano 395-0808, Japan}
\email{takehiro.kume@natsume-optics.co.jp}  

\author[orcid=0000-0002-2409-3956]{Yusuke Matsuzawa} 
\affiliation{Natsume Optical Corporation, 3461 Kamichyaya, Kanae, Iida, Nagano 395-0808, Japan}
\email{yusuke.matsuzawa@natsume-optics.co.jp}  

\author[orcid=0009-0000-9348-9164]{Yoichi Imamura} 
\affiliation{Natsume Optical Corporation, 3461 Kamichyaya, Kanae, Iida, Nagano 395-0808, Japan}
\email{yoichi.imamura@natsume-optics.co.jp}  

\author{Takahiro Saito} 
\affiliation{Natsume Optical Corporation, 3461 Kamichyaya, Kanae, Iida, Nagano 395-0808, Japan}
\email{takahiro.saito@natsume-optics.co.jp}  

\author[orcid=0000-0003-2231-3502]{Kentaro Hiraguri} 
\affiliation{Natsume Optical Corporation, 3461 Kamichyaya, Kanae, Iida, Nagano 395-0808, Japan}
\email{kentaro.hiraguri@natsume-optics.co.jp}  

\author[orcid=0000-0002-4589-0460]{Hirokazu Hashizume} 
\affiliation{Natsume Optical Corporation, 3461 Kamichyaya, Kanae, Iida, Nagano 395-0808, Japan}
\email{hirokazu.hashizume@natsume-optics.co.jp}  

\author[orcid=0000-0003-4623-6951]{Hidekazu Mimura} 
\affiliation{RIKEN/ SPring-8, 1-1-1, Kouto, Sayo-cho, Sayo-gun, Hyogo 679-5198, Japan}
\affiliation{Research Center for Advanced Science and Technology, The University of Tokyo, 4-6-1 Komaba, Meguro-ku, Tokyo 153-8904, Japan}
\email{mimura@upm.rcast.u-tokyo.ac.jp}




\begin{abstract}

We have developed high angular resolution electroformed nickel X-ray telescope mirrors by extending precision mirror-fabrication techniques originally developed for synchrotron X-ray microscopy to space astronomy. 
Our previous study identified axial figure errors of a replicated mirror as one of the factors limiting its X-ray angular resolution. 
In the present study, we fabricated a new mirror using a substantially improved mandrel and directly compared the corresponding two-dimensional figure errors of the mandrels and replicated inner mirror surfaces for both the previous and new mandrel--mirror pairs. 
This comparison enables us to investigate how the relative contributions to the final mirror figure errors changed following the improvement of the mandrel accuracy. For the previous mirror, the mandrel and mirror showed a strong spatial correlation in their figure errors, indicating that a substantial component of the mirror figure pattern was inherited from the mandrel. 
In contrast, after substantial improvement of the mandrel figure, the new mirror exhibited a weaker mandrel--mirror spatial correlation, while the residual figure deviations arising between the mandrel and final mirror became comparable in magnitude to the residual mandrel figure errors. 
The improved mandrel reduced the central 68\% widths, W$_{68}$, of the axial slope error distribution from 3.3 to 1.6 arcsec for the paraboloidal surface and from 5.0 to 1.1 arcsec for the hyperboloidal surface. 
Correspondingly, the mirror values improved from 4.6 to 2.4 arcsec and from 12.4 to 2.0 arcsec, respectively. 
X-ray measurements at 12 keV yielded local half power diameters (HPDs) of approximately 7--17 arcsec and a full-aperture HPD of 8.5 arcsec. 
The newly fabricated mirror was subsequently adopted for the soft X-ray telescope mirror of the fifth flight of the Focusing Optics X-ray Solar Imager (FOXSI-5) sounding rocket experiment and was successfully flown in May 2026.

\end{abstract}

\keywords{\uat{High Energy astrophysics}{739} --- \uat{Astronomical optics}{88} --- \uat{High angular resolution}{2167} --- \uat{Solar physics}{1476} --- \uat{X-ray astronomy}{1810}}

\section{Introduction} \label{sec:Intro}

High-angular-resolution X-ray imaging is essential for investigating high-energy phenomena in the Universe. 
Grazing-incidence X-ray optics based on Wolter-type geometries have therefore played a central role in astronomical X-ray observations. 
Various fabrication approaches have been developed to meet the demanding requirements on the figure accuracy, surface quality, effective area, and structural properties of astronomical X-ray telescope mirrors \citep[e.g.,][]{Atkins2022book}.
Among various fabrication techniques, electroforming replication has been successfully employed for a wide range of X-ray instruments, including satellite observatories such as XMM-Newton, eROSITA aboard Spectrum-Roentgen-Gamma (SRG), and IXPE \citep{Aschenbach2002spie,Predehl2021a&a,Ramsey2022jatis}, as well as balloon-borne and sounding rocket X-ray experiments \citep[e.g.,][]{Ramsey1999spie,Krucker2013spie}.
Electroforming replication is particularly attractive for fabricating double reflection, full-circumference monolithic X-ray mirror shells, in which the primary and secondary reflecting surfaces can be replicated as a single coaxial structure. 
Such a configuration provides high structural stiffness and simplifies mirror alignment and assembly \citep{Pareschi2005spie,Kilaru2019jatis}. 
In addition, once a precision mandrel has been fabricated, it can be reused to produce multiple nominally identical shells, offering potential cost advantages when multiple mirrors or mirror modules are required \citep{Kilaru2019jatis,Singam2024spie}.
In this approach, the surface figure of a precisely fabricated mandrel is replicated onto the inner reflecting surface of an electroformed shell, making both the mandrel figure and replication fidelity important factors in determining the imaging performance.
Continuous efforts have been made to improve the angular resolution of electroformed X-ray optics. 
Previous studies have investigated improvements in mandrel fabrication and polishing, control of the electroforming and separation processes, and correction of the resulting mirror figure \citep[e.g.,][]{Pareschi2005spie,Gubarev2010nims,Kilaru2019jatis}. 
These studies have demonstrated that the final imaging performance depends not only on the accuracy of the mandrel but also on how faithfully its figure is transferred to the replicated mirror. 
Evaluating both the mandrel and the resulting mirror is therefore important for identifying the dominant sources of figure error and further improving electroformed X-ray optics.

Our approach is based on a precision electroforming technique originally developed for millimeter-aperture, ultra-short-focal-length X-ray mirrors for synchrotron X-ray microscopy \citep{Kume2019rsi,Mimura2018rsi,Takeo2020apphl,Takeo2020apphl2}. 
Unlike conventional electroforming processes typically performed at elevated temperatures with bath agitation, our process employs room-temperature, non-agitated deposition. 
A quartz-glass mandrel with a low thermal-expansion coefficient is used to suppress thermally induced deformation, while the electroforming conditions are optimized for highly accurate transfer of the mandrel surface \citep{Kume2019rsi,Mimura2018rsi}. 
This technique was subsequently scaled from millimeter-aperture mirrors for synchrotron X-ray microscopy to a 60-mm-diameter, full-circumference, double-reflection Wolter-I mirror for astronomical applications \citep{Yamaguchi2020spie,Ito2021spieproc,Yamaguchi2023rsi,Fujii2026pasp}.
This electroforming technique was subsequently applied to X-ray telescope mirrors developed for the fourth flight of the Focusing Optics X-ray Solar Imager (FOXSI-4) sounding rocket experiment, demonstrating its applicability to flight X-ray optics \citep{Buitrago-Casas2021spie,Ampuku2024spie}. 
In our previous study, \citet{Fujii2026pasp} reported detailed X-ray characterization of a 60-mm-diameter electroformed mirror and its mirror module assembly (MMA). 
The mirror achieved a sub-arcsecond PSF core and a half power diameter (HPD) of approximately 14 arcsec under full-aperture illumination at 12 keV and was adopted as one of the hard X-ray optics for FOXSI-4.

Despite these advances, the origin of the remaining figure errors had not been fully clarified. 
In \citet{Fujii2026pasp}, local imaging performance measured by spot scans was compared with the axial slope error distribution derived from the mirror figure data in the corresponding illuminated regions, revealing a positive correlation between them. 
However, the spatial correspondence between the figure errors of the mandrel and the replicated inner mirror surface was not investigated. 
Despite the challenges associated with characterizing the inner reflecting surface, such a spatially resolved comparison is particularly important for X-ray telescope mirrors fabricated by replication techniques, because it allows figure errors inherited from the mandrel to be distinguished from subsequent figure changes introduced during the replication process.

In this study, we fabricated a new electroformed Wolter-I telescope mirror using a mandrel with substantially improved figure accuracy. 
Using the inner-surface metrology established in our previous development of mirrors for synchrotron X-ray microscopy \citep{Kume2019rsi}, we directly compare corresponding two-dimensional figure-error distributions of the mandrel and the replicated inner mirror surface over their full circumferences, thereby investigating the spatial fidelity of figure transfer and how improvement of the mandrel figure is reflected in the final mirror and its X-ray imaging performance. 
The imaging performance of the newly fabricated mirror was evaluated using synchrotron X-rays at SPring-8 through local and full-aperture measurements. 
Based on its demonstrated imaging performance, the mirror was adopted for the soft X-ray optics of the fifth flight of the Focusing Optics X-ray Solar Imager (FOXSI-5) sounding rocket experiment \citep{Yoshida2026spie}.

Throughout this paper, all data analyses were performed using Python 3.13.5 and standard scientific software packages, including Astropy 7.1.0 \citep{astropy:2013, astropy:2018, astropy:2022}, NumPy 2.2.1 \citep{harris2020array}, SciPy 1.16.0 \citep{2020SciPy-NMeth},
Matplotlib 3.10.3 \citep{Hunter:2007}, and pandas 2.3.1 \citep{reback2020pandas}.

\section{Fabrication and Figure Characterization} \label{chap_sample}
\subsection{Fabrication of the Newly-Fabricated Electroformed X-ray Mirror with Improved Mandrel} \label{sub_sample}

The X-ray mirror developed in this study has the same basic optical design as that reported in \citet{Fujii2026pasp}, which is a double reflection,
full-circumference monolithic Wolter-I mirror incorporating both the primary paraboloidal and secondary hyperboloidal reflecting surfaces. 
The mirror has a total axial length of 220 mm, an aperture diameter of 60 mm, and a focal length of 2000 mm. 
The nominal grazing-incidence angle is approximately 0.21 deg. The mirror is made of electroformed nickel without an additional reflective coating.
The basic fabrication and replication procedures follow those established in \citet{Yamaguchi2020spie,Yamaguchi2023rsi} and \citet{Fujii2026pasp} and are therefore only briefly described here. 
A precision-polished mandrel serves as the master surface for electroforming the nickel mirror, transferring its surface figure to the inner reflecting surface. 
\citet{Fujii2026pasp} demonstrated that the local X-ray imaging performance of the previous mirror was related to its axial figure errors, indicating the need for improved figure accuracy. 
Accordingly, the mandrel used in this study was refined through repeated corrective polishing and figure measurements, with particular emphasis on reducing axial figure errors.
At each iteration, the amount and spatial distribution of material removal were determined from the measured figure errors, allowing the spatial pattern of the residual figure errors to evolve during successive polishing cycles.
%


\subsection{Mandrel and Mirror Figure-error Measurements} \label{sec:figerr}
\begin{figure*}[h!]
  \centering
  \includegraphics[width=0.9\linewidth]{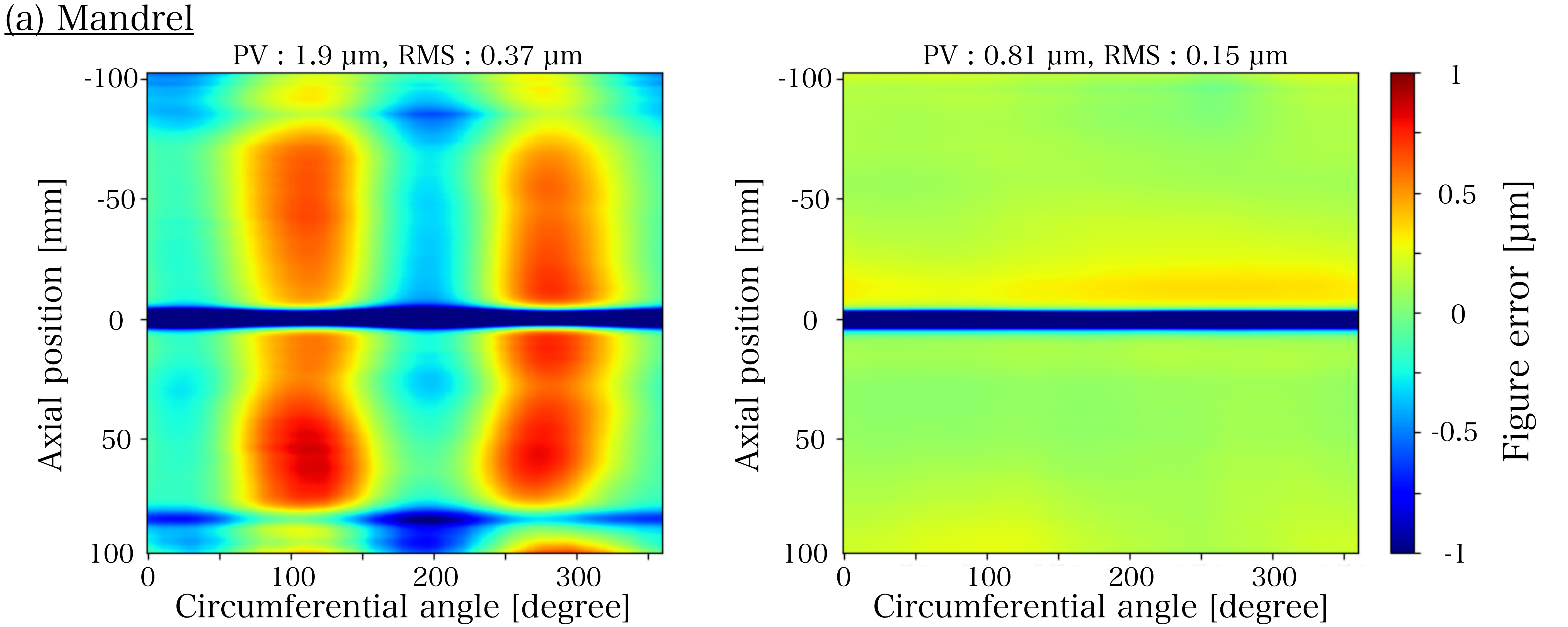}
  \vspace{2mm}
  \includegraphics[width=0.9\linewidth]{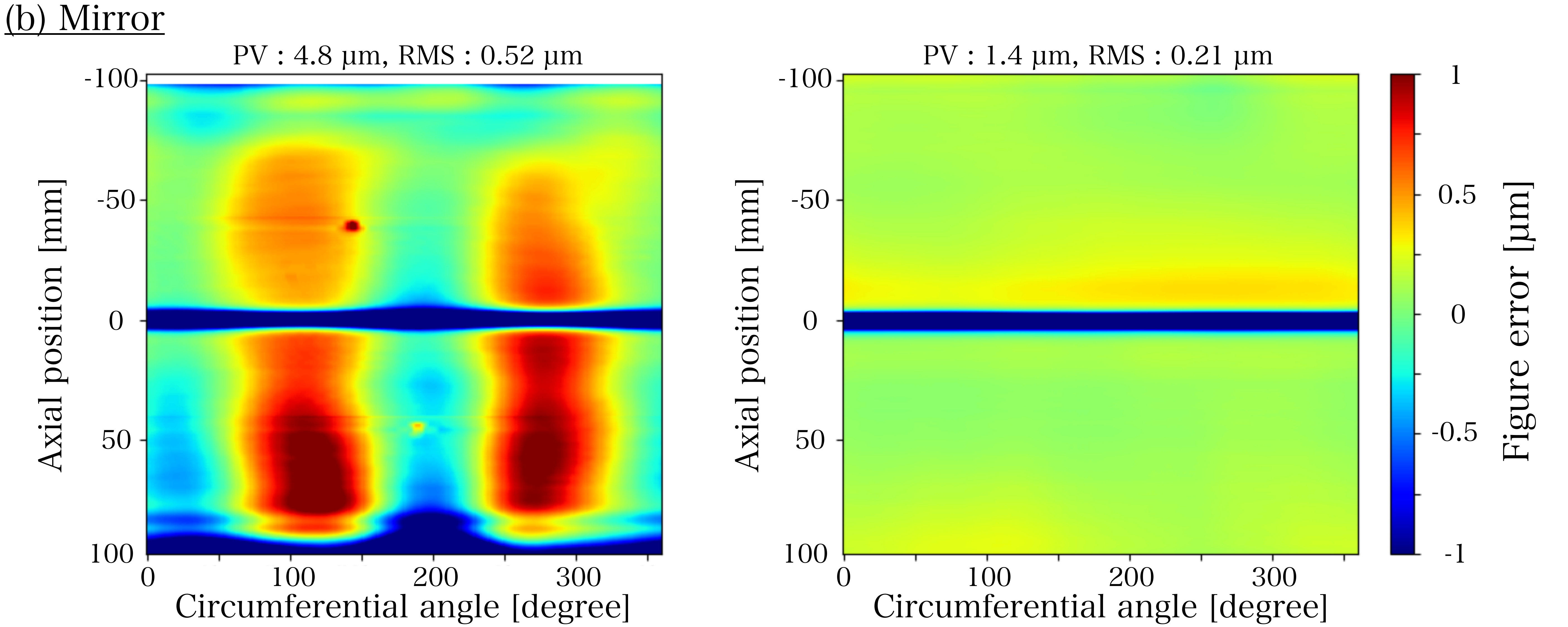}
  \vspace{2mm}
  \includegraphics[width=0.9\linewidth]{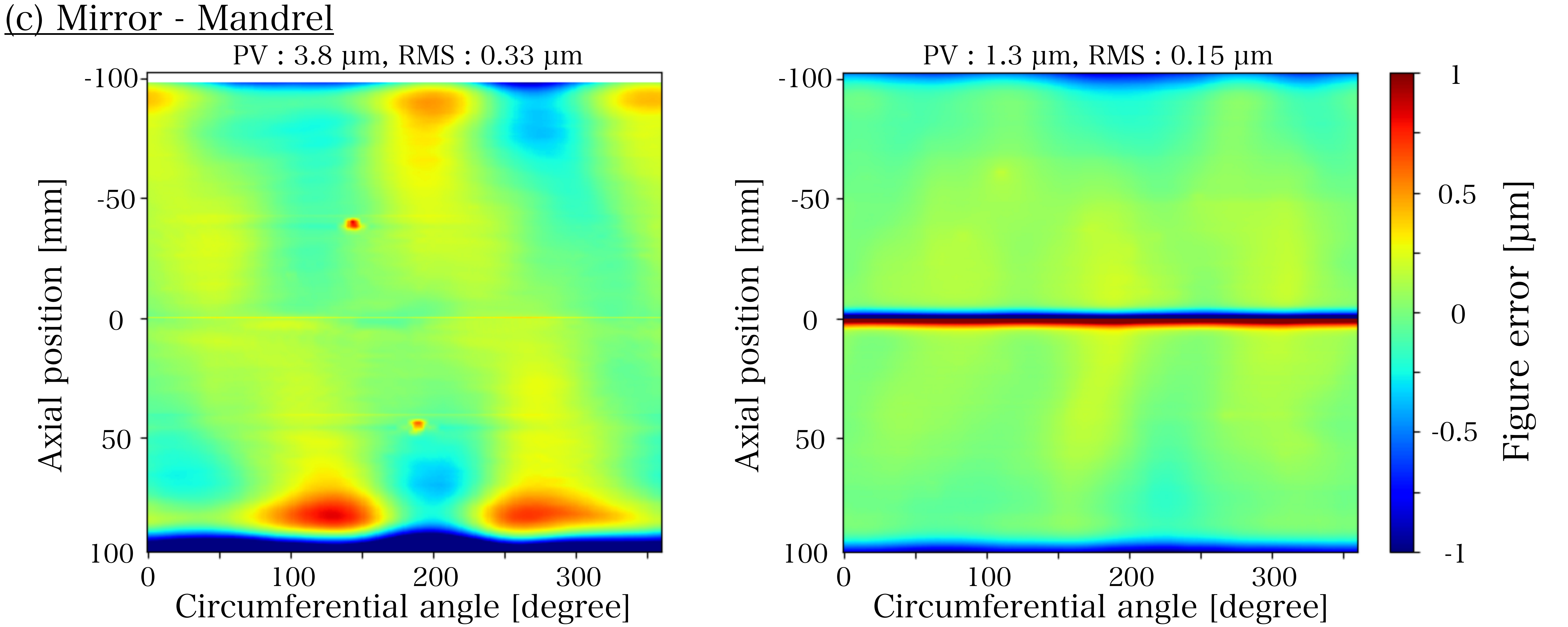}
  \caption{Two-dimensional figure-error maps of the previous and improved mandrels, their corresponding replicated mirrors, and the mandrel-to-mirror figure differences. 
  Panels (a), (b), and (c) show the mandrels, replicated mirrors, and mandrel-to-mirror figure differences, respectively.
  In each panel, the left and right maps show the results for the previous and new mandrel--mirror pairs, respectively. 
  The mandrel-to-mirror figure difference was obtained by subtracting the mandrel figure error from the corresponding mirror figure error at each spatial position.
  The horizontal and vertical axes represent the circumferential angle and axial position, respectively.
  The figure error represents the radial deviation from the ideal Wolter-I surface, measured perpendicular to the optical axis.  
  The PV and RMS values shown in each map were calculated after excluding the region within $\pm5$ mm of the boundary between the paraboloidal and hyperboloidal surfaces, which contains the discontinuous and unpolished boundary.
  The same color scale is used for all maps to facilitate direct comparison.}
  \label{fig:figure_2d-figureerror-map}
\end{figure*}

The figure errors of the mandrel and replicated mirror were measured over their full circumferences. The figure error is defined as the radial deviation of the measured surface from the ideal Wolter-I geometry, measured perpendicular to the optical axis. 
Measurements were performed with a pitch of 50~$\mu$m in the axial direction and an angular pitch of 1~degree in the circumferential direction, providing two-dimensional figure-error maps over the full 360~degree circumference. 
This spatially resolved dataset enables the mandrel and replicated inner mirror surface to be compared at corresponding positions over the full circumference, which is important for quantitatively identifying the origins of figure errors that degrade the X-ray angular resolution.
%
%
The measurement repeatability, evaluated from two independent measurements, was approximately 10~nm (RMS). 
%
%
The boundary between the paraboloidal and hyperboloidal surfaces is discontinuous and includes an unpolished region. Therefore, as in \citet{Fujii2026pasp}, data within $\pm5$~mm of the boundary were excluded from the quantitative analyses. 
This exclusion was applied to the calculation of the peak-to-valley (PV) and root-mean-square (RMS) values, the axial slope error distributions, and the two-dimensional correlation analyses described below.
Figure~\ref{fig:figure_2d-figureerror-map} shows the two-dimensional figure-error maps of the previous and improved mandrels and their corresponding replicated mirrors. 
To characterize the change in figure between each mandrel and the corresponding replicated mirror, we also calculated the mandrel-to-mirror figure difference by subtracting the mandrel figure error from the mirror figure error at each spatial position. 
The corresponding two-dimensional maps are also shown in Figure~\ref{fig:figure_2d-figureerror-map}. Because the mandrel-to-mirror figure difference can contain measurement and spatial-registration uncertainties in addition to figure changes arising during the fabrication process, we use the term ``mandrel-to-mirror figure difference'' throughout this paper rather than attributing it solely to the electroforming process.
A qualitative comparison of the maps shows smaller figure-error variations for the improved mandrel and newly fabricated mirror than for their previous counterparts. 
Quantitatively, the PV and RMS values of the mandrel decreased from 1.9 and 0.37 $\mu$m to 0.81 and 0.15 $\mu$m, respectively. 
The corresponding values for the replicated mirror decreased from 4.8 and 0.52 $\mu$m to 1.4 and 0.21 $\mu$m. 
The mandrel-to-mirror figure difference also showed smaller variations, with the PV and RMS values decreasing from 3.8 and 0.33 $\mu$m to 1.3 and 0.15 $\mu$m, respectively. 
These improvements are further quantitatively characterized using the axial slope error distributions and spatial correlations in the following subsections.
\subsection{Mandrel and Mirror Axial Slope Error Distributions}

\begin{figure*}[h!]
  \centering
  \includegraphics[width=0.9\linewidth]{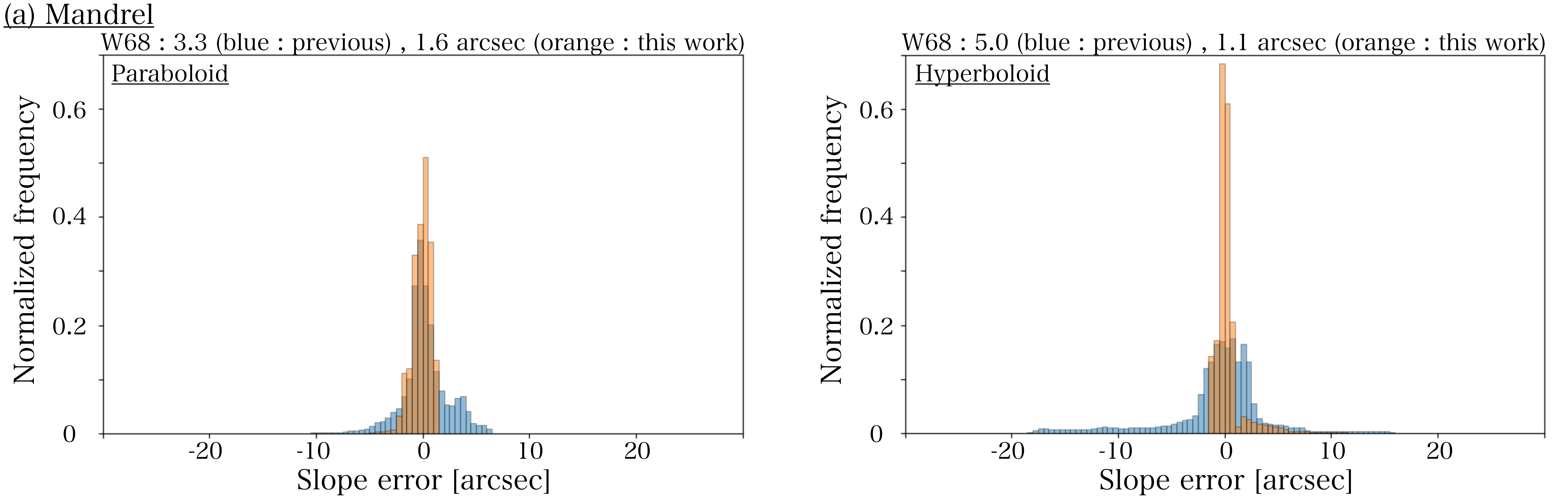}
  \vspace{2mm}
  \includegraphics[width=0.9\linewidth]{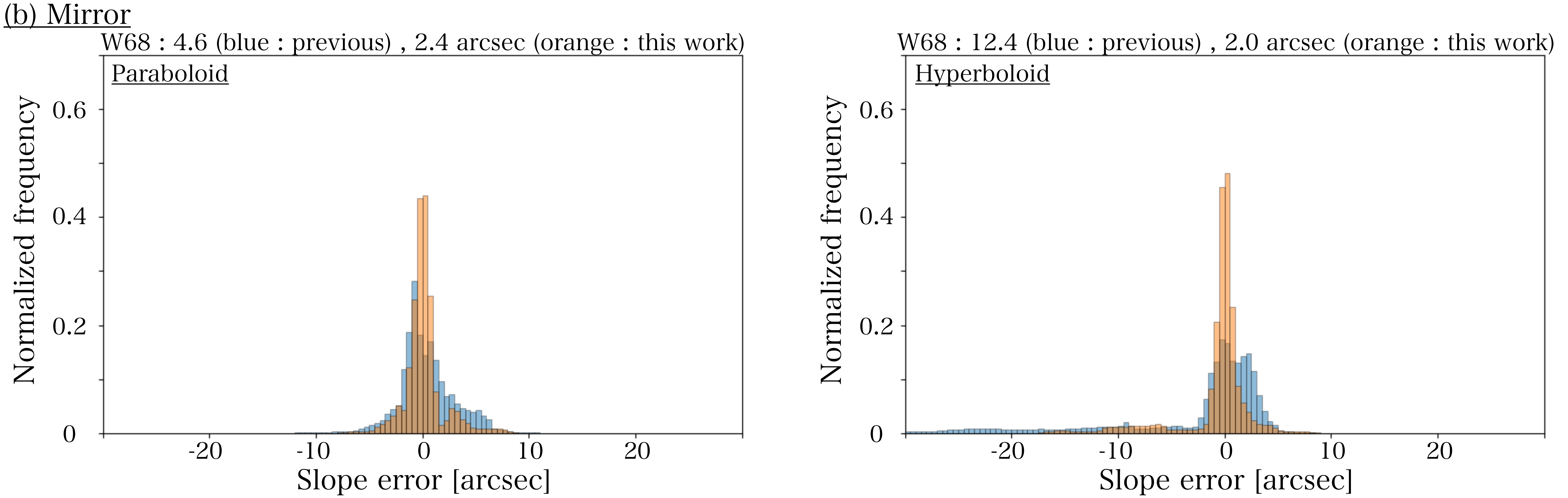}
  \vspace{2mm}
  \includegraphics[width=0.9\linewidth]{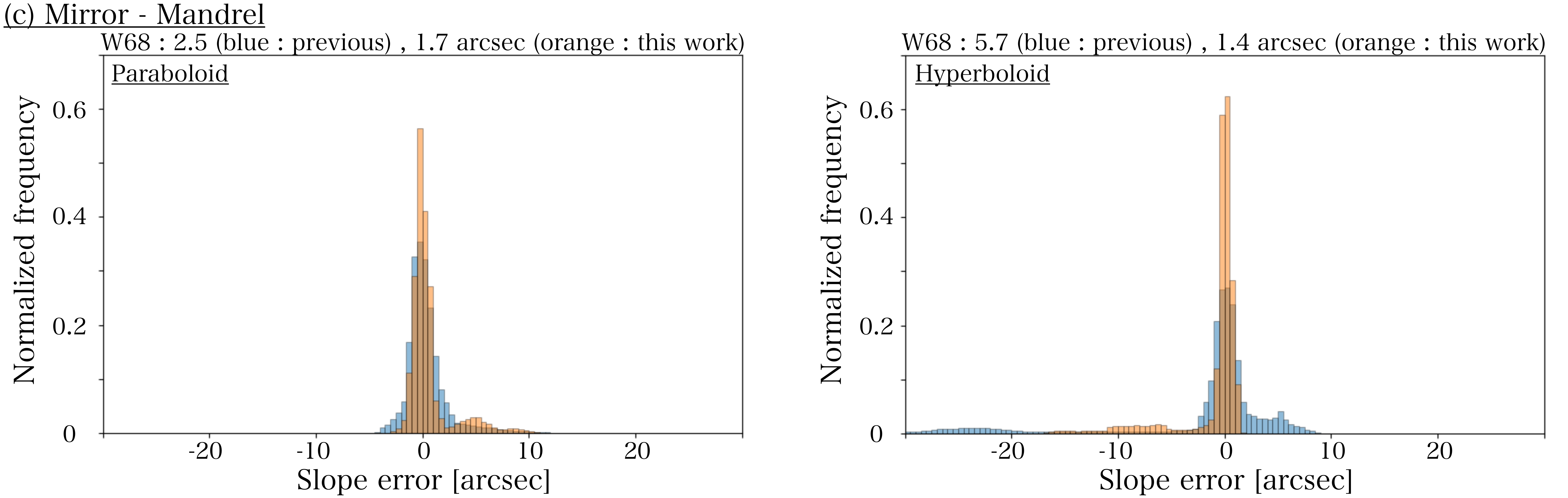}
  \caption{Axial slope error distributions derived from the two-dimensional figure-error maps shown in Figure~\ref{fig:figure_2d-figureerror-map}. 
  The left and right panels show the distributions for the paraboloidal and hyperboloidal surfaces, respectively. 
  Panels (a), (b), and (c) show the axial slope error distributions of the mandrels and replicated mirrors and the corresponding mandrel-to-mirror figure differences, respectively.
  In each panel, the previous results and those obtained in this work are compared.
  For comparison of the spreads of the distributions independently of their absolute offsets, the median of each distribution was subtracted from the individual slope error values.
  Each histogram was normalized by the total number of data points.
  The spread of each distribution is characterized by $W_{68}$, defined as the difference between the 84th and 16th percentiles.
  The corresponding $W_{68}$ values for the previous results and those obtained in this work are shown in each panel.
  The region within $\pm5$ mm of the boundary between the paraboloidal and hyperboloidal surfaces was excluded from the analysis.}
  \label{fig:figure_slopeerrorhist}
\end{figure*}

To quantitatively characterize the figure accuracy relevant to the X-ray imaging performance, axial slope errors were derived from the measured 
figure-error data separately for the paraboloidal and hyperboloidal surfaces.
The region within $\pm 5$ mm of the boundary between the two surfaces was excluded from the analysis. 
As in \citet{Fujii2026pasp}, the median value was subtracted from each axial slope error distribution so that the median was centered at zero. 
Figure~\ref{fig:figure_slopeerrorhist} shows the resulting axial slope error distributions derived from the figure-error maps shown in Figure~\ref{fig:figure_2d-figureerror-map}. 
The distributions, particularly those for the previous mirror, are asymmetric and exhibit extended tails. 
We therefore characterize the spread of each distribution using percentile-based quantities rather than assuming a Gaussian distribution. 
We define $W_{68}$ as the difference between the 84th and 16th percentiles.
For the previous mandrel, $W_{68}$ was 3.3 and 5.0 arcsec for the paraboloidal and hyperboloidal surfaces, respectively. 
For the improved mandrel, these values decreased to 1.6 and 1.1 arcsec, corresponding to reductions of approximately 52\% and 78\%, respectively. 
The improvement was also observed for the replicated mirror. 
The $W_{68}$ values decreased from 4.6 to 2.4 arcsec for the paraboloidal surface and from 12.4 to 2.0 arcsec for the hyperboloidal surface, corresponding to reductions of approximately 48\% and 84\%, respectively. 
The mandrel-to-mirror figure differences also showed narrower distributions, with $W_{68}$ decreasing from 2.5 to 1.7 arcsec for the paraboloidal surface and from 5.7 to 1.4 arcsec for the hyperboloidal surface, corresponding to reductions of approximately 32\% and 75\%, respectively.
Thus, smaller $W_{68}$ values were obtained for the improved mandrel, the newly fabricated mirror, and the corresponding mandrel-to-mirror figure 
difference for both reflecting surfaces. 
Notably, the previous mirror exhibited a large difference in $W_{68}$ between the paraboloidal and hyperboloidal surfaces, whereas the newly fabricated mirror yielded comparable values of 2.4 and 2.0 arcsec, respectively. 
At the same time, some of the slope error distributions of both the previous and newly fabricated mirrors exhibit extended tails that are not apparent in the corresponding mandrel distributions. 
For the new mirror, the residual $W_{68}$ values of the improved mandrel, 1.6 and 1.1 arcsec for the paraboloidal and hyperboloidal surfaces, respectively, are comparable to those of the corresponding mandrel-to-mirror figure differences, 1.7 and 1.4 arcsec. 
The implications of these residual figure differences for further improvement of the mirror figure are discussed in Section~\ref{sec:discussion}.

\subsection{Spatial Correlation of the Mandrel and Mirror Figure Errors} \label{subsec:figerr-correlations}
\begin{figure*}[h!]
  \centering
  \includegraphics[width=0.9\linewidth]{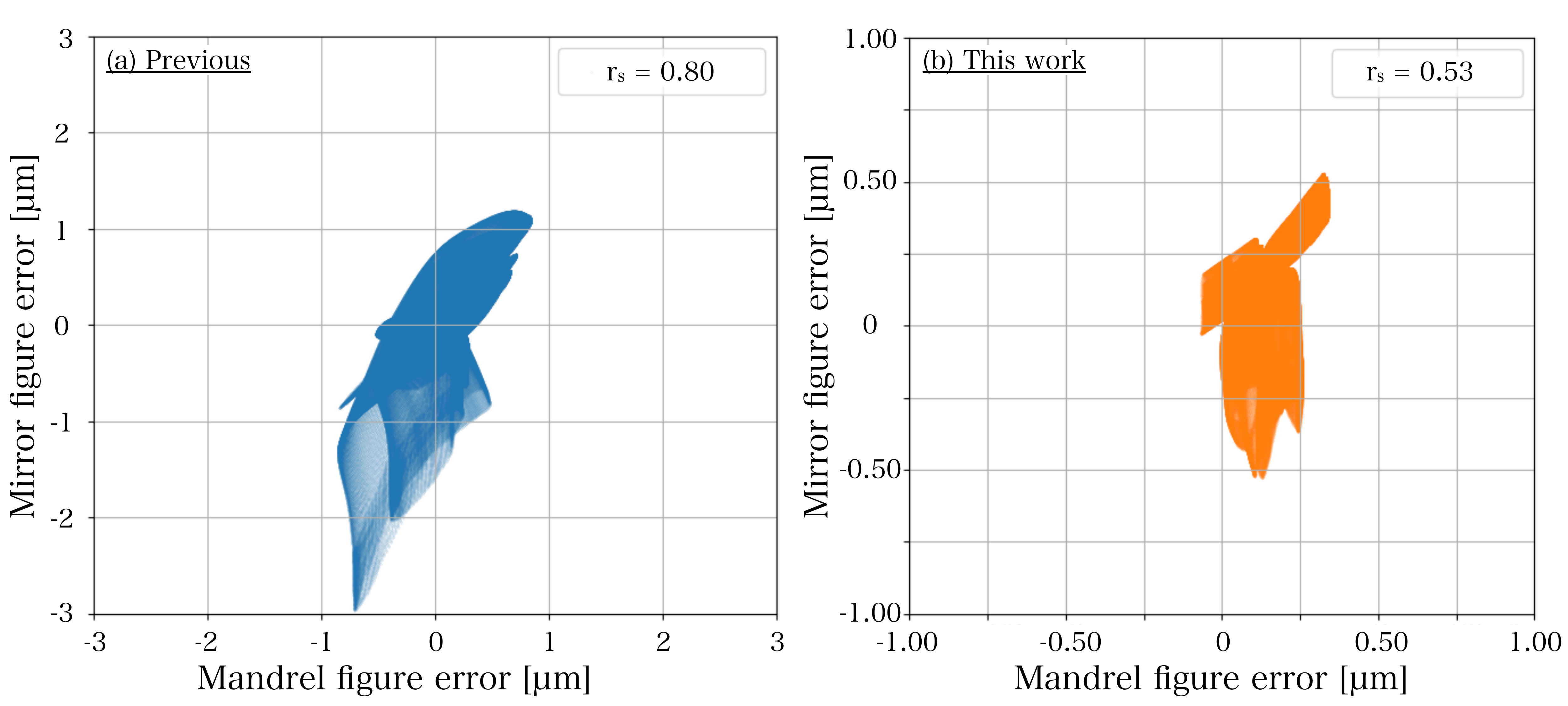}
  \vspace{2mm}
  \caption{Spatial correlations between the two-dimensional figure errors of the mandrels and the corresponding replicated mirrors. Panels (a) and (b) show the previous results and those obtained in this work, respectively. Each point represents the figure errors of the mandrel and mirror at the same spatial position after applying a 10 mm running average in the axial direction. The region within $\pm5$ mm of the boundary between the paraboloidal and hyperboloidal surfaces was excluded from the analysis. The Spearman's rank correlation coefficient, $r_{\rm s}$, is shown in each panel.
  %
  %
  %
  %
  }
  \label{fig:figure_2dfedcorrelation}
\end{figure*}

To investigate how the mandrel figure is spatially reflected in the replicated mirror, we performed a pixel-by-pixel correlation analysis of the corresponding two-dimensional figure-error maps.
%
Consistent with the spatial scale adopted in the figure-error analysis of \citet{Fujii2026pasp}, a 10 mm running average was applied in the axial 
direction before calculating the spatial correlations. 
As in the preceding analyses, the $\pm 5$ mm region around the boundary between the two reflecting surfaces was excluded. 
Figure~\ref{fig:figure_2dfedcorrelation} compares the spatial correlations between the mandrel and mirror figure errors for the previous and newly fabricated mirrors.
For the previous results, the Spearman's rank correlation coefficient between the mandrel and mirror figure errors was $r_{\rm s}=0.80$, indicating a strong spatial correspondence between their figure-error patterns. 
In contrast, for the same mandrel--mirror pair, the correlation between the mandrel and mandrel-to-mirror figure difference was only $r_{\rm s}=0.11$, indicating little spatial correspondence between these quantities. 
For the newly fabricated mirror, the mandrel--mirror correlation decreased to $r_{\rm s}=0.53$, whereas the correlation between the mandrel and mandrel-to-mirror figure difference remained negligible, $r_{\rm s}=0.08$.

As seen in Figure~\ref{fig:figure_2dfedcorrelation}, both the previous and newly fabricated mirrors exhibit populations of points that deviate from the main mandrel--mirror relation. 
Tracing these points back to their spatial locations showed that the prominent branches are associated mainly with regions near the axial ends. 
Such branches are present in both mirrors, although their prominence differs between them.
%
%
%
These results indicate that the spatial pattern of the previous mirror figure strongly reflected that of the mandrel, while the mandrel-to-mirror figure difference was not systematically related to the local mandrel figure in either generation of the mirrors.
Because the two-dimensional comparison requires spatial registration of independently measured maps, we evaluated the sensitivity of the correlation coefficients to possible circumferential registration errors. 
One map was intentionally displaced by $\pm$1 sampling point, corresponding to $\pm$1 degree in the circumferential direction, and the correlation coefficients were recalculated. 
For all correlations examined, the changes in the Spearman’s rank correlation coefficients were less than 0.01. 
The effect of circumferential misregistration at the level of one sampling point is therefore negligible and does not affect any of the conclusions drawn from the spatial-correlation analysis.
%
%
%
%

%
%
%
\section{X-ray Imaging Performance} \label{sec:results}
\subsection{Experimental Setup and Measurement Procedure}\label{subsec:setup}

The X-ray imaging performance of the new mirror was evaluated at 12 keV using essentially the same experimental configurations as those described by \citet{Fujii2026pasp}, except for the CMOS camera. 
In the present measurements, a Hamamatsu Photonics ORCA-Quest qCMOS camera (C15550-20UP), with $4096 \times 2304$ pixels and a pixel size of 4.6 $\mu$m, was used. 
The smaller pixel size than the 6.5 $\mu$m pixels used in \citet{Fujii2026pasp} provided finer sampling of the focal-plane image.
Two complementary measurements were performed: local spot scan measurements using a narrow X-ray beam and full-aperture measurements using the HBX-KLAEES system.
In the spot scan measurements, a narrow X-ray beam was sequentially irradiated onto the mirror aperture. 
A focal-plane image was obtained at each illuminated position, and the local angular resolution was evaluated in terms of the full width at half maximum (FWHM) and HPD. 
The spot scan measurements therefore characterize the spatial variation of the imaging performance over the mirror.
For the full-aperture measurement, the entire mirror aperture was illuminated simultaneously with an expanded 12 keV X-ray beam produced by the HBX-KLAEES system. 
The angular resolution was derived from the encircled-energy distribution of the measured focal-plane image and provides a direct measurement of the integrated imaging performance of the complete aperture.
Because the basic mirror design and measurement configurations were essentially unchanged from those used by \citet{Fujii2026pasp}, the present measurements provide a direct comparison of the X-ray imaging performance before and after the improvement of the mandrel figure.

\subsection{Local X-ray Imaging Performance}\label{subsec:local}
\begin{figure*}[h!]
  \centering
  \includegraphics[width=0.9\linewidth]{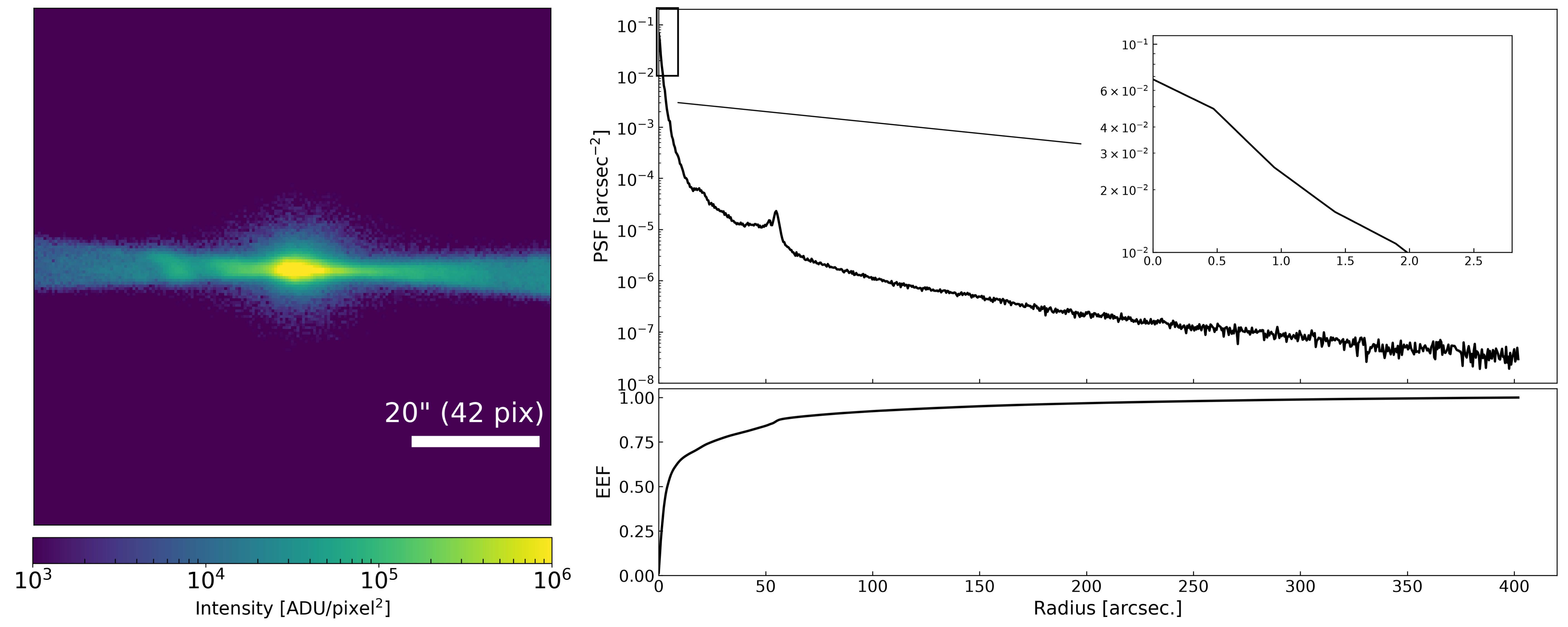}
  \vspace{2mm}
  \includegraphics[width=0.9\linewidth]{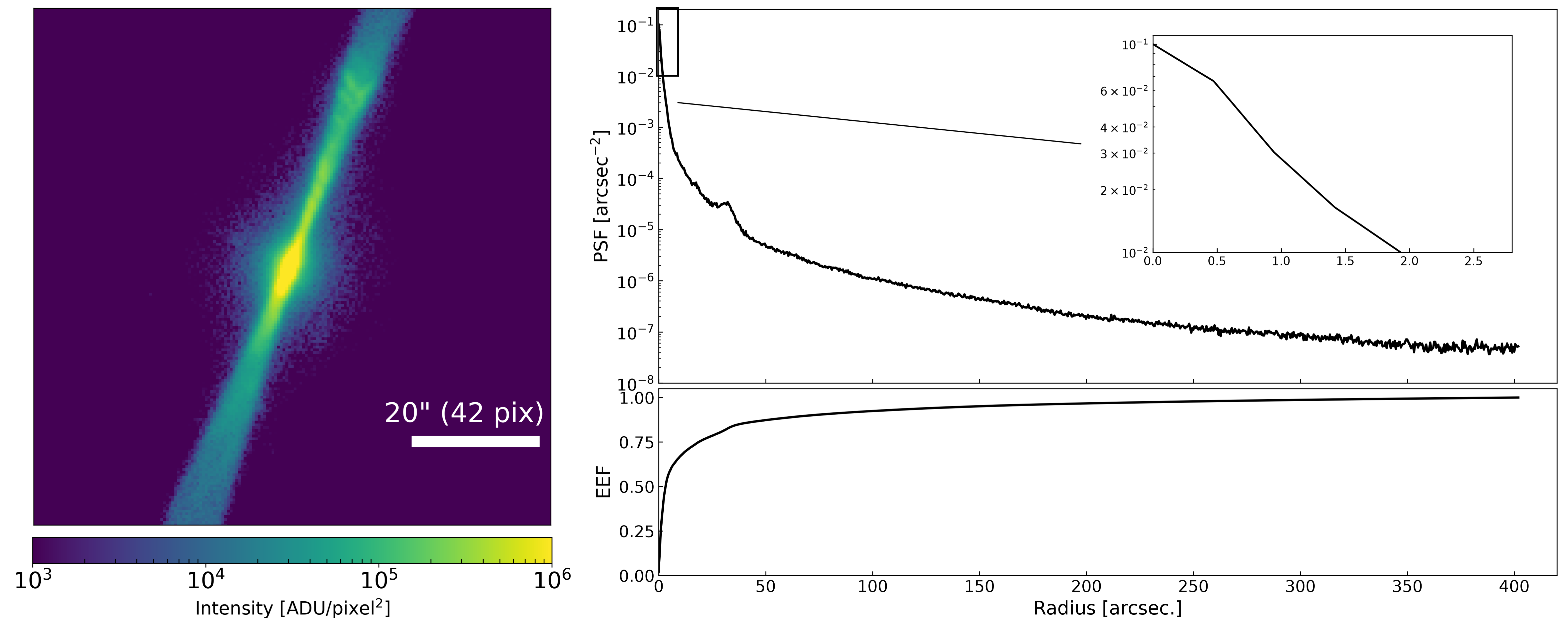}
  \caption{Representative results of the local spot scan measurements at 12~keV.
  The upper and lower rows show the results at a position yielding a typical HPD and at one of the positions yielding the smallest HPD, respectively.
  The typical position was selected as the measurement position with an HPD closest to the median value among the 40 measured positions.
  For each measurement position, the focal-plane image is shown on the left, while the corresponding PSF and EEF are shown in the upper-right and lower-right panels, respectively.}
  \label{fig:figure_spotscan}
\end{figure*}

After the alignment and focusing procedures described in \citet{Fujii2026pasp}, the acquired spot scan images were analyzed to evaluate the local imaging performance
of the mirror. 
For background correction, a dark frame was acquired by closing the upstream beam shutter under the same conditions and was subtracted from each illuminated image. 
Because of the relatively low photon statistics in the spot scan measurements, short-term variations of the background were also evaluated using source-free regions within the field of view and incorporated into the background estimation.
The background variations occurred on timescales of approximately 7--9 min, as well as over the full
measurement period of approximately 50 min. The background correction factors ranged from 0.99980 to 1.00017 relative to the constant dark frame, corresponding to deviations of at most 0.020\%. These variations may be associated with small fluctuations in the detector readout offset or associated electronics. Although these corrections were small, they were important for accurately characterizing the faint wings of the PSF over a dynamic range of approximately six orders of magnitude.
The angular resolution was evaluated using the FWHM and HPD. 
The PSF was obtained by extracting the radial intensity profile centered at the maximum pixel value, and the encircled energy function (EEF) was calculated by integrating and normalizing this profile. 
The HPD was determined as twice the radius corresponding to an EEF of 0.5. 
During the analysis, the extraction radius was carefully selected to exclude direct X-rays that bypassed the mirror and stray light produced by single reflections from either the paraboloidal or hyperboloidal surface.
Direct X-rays were blocked by a mask consisting of 4-mm-thick aluminum and 2-mm-thick lead, with a diameter of approximately 30 mm, positioned immediately upstream of the mirror (see Figure 3 of \citet{Fujii2026pasp}). The mask was sufficiently thick to suppress higher-order X-rays. For on-axis illumination, single-reflection stray light is theoretically expected at a radial distance of approximately 5 mm (1100 pixels) from the optical axis, outside the extraction radius of approximately 3.7 mm (800 pixels).
Table~\ref{table:spotscan_hpd} summarizes the HPD measured at each of the 40 spot scan positions at 12~keV, together with the corresponding slope error spreads of the paraboloidal and hyperboloidal surfaces. 
Following \citet{Fujii2026pasp}, the median absolute deviation from the baseline was adopted as the measure of the local slope-error spread.
The measured HPDs ranged from 6.6 to 17.1~arcsec. 
Most of the illuminated regions showed HPDs of approximately 7--9~arcsec, while larger values exceeding 10~arcsec were obtained only in limited regions of the mirror. 
The smallest measured HPD was 6.6~arcsec.
The measured FWHM was 1.9~arcsec at all 40 spot scan positions. 
As discussed in \citet{Fujii2026pasp}, the FWHM is limited by the spatial sampling of the focal-plane detector. 
In the present measurements, the smaller pixel size of 4.6~$\mu$m reduced this sampling limit compared with the 6.5~$\mu$m pixels used in \citet{Fujii2026pasp}; nevertheless, the uniform measured value of 1.9~arcsec indicates that the intrinsic PSF core was not fully resolved. 
The intrinsic FWHM may therefore be smaller than 1.9~arcsec. 
In contrast, the HPD, which includes the broader components of the PSF, exhibits measurable spatial variation over the mirror aperture. 
We therefore use the HPD as the primary metric for characterizing the spatial dependence of the local imaging performance.
Figure~\ref{fig:figure_spotscan} shows representative spot scan results at 12~keV.
The upper row shows the measurement position with an HPD closest to the median of the 40 measurements, while the lower row shows a position yielding the smallest HPD. 
For each position, the focal-plane image, PSF, and EEF are shown. 
These measurements demonstrate a narrow central PSF core together with broader components that determine the HPD, including a weak shoulder-like structure at several tens of arcseconds in some measurements.

\begin{table}[htbp]
\centering
\hfill
\begin{minipage}{0.9\textwidth}
\centering
\begin{tabular}{lcc|lcc|lcc} \hline
Position & HPD & slope error\footnote{The slope-error values represent the median absolute deviations from the baseline for the primary and secondary reflecting surfaces, respectively, listed to the left and right of the slash.\label{fot:a}} & Position & HPD & slope error\footref{fot:a} & Position & HPD & slope error\footref{fot:a} \\
{[deg.]} & {[arcsec.]} & {[arcsec.]} & {[deg.]} & {[arcsec.]} & {[arcsec.]} & {[deg.]} & {[arcsec.]} & {[arcsec.]} \\ \hline
  -5 - 4 & 7.6 & 0.91/0.73 & 125 - 132 & 7.6 & 0.72/0.66 & 245 - 255 & 6.6 & 0.46/0.35 \\
  4 - 14 & 7.6 & 0.92/0.56 & 132 - 138 & 7.6 & 0.54/0.65 & 255 - 265 & 6.6 & 0.45/0.38 \\
  14 - 24 & 6.6 & 0.98/0.48 & 138 - 144 & 8.5 & 0.40/0.67 & 265 - 274 & 6.6 & 0.54/0.44 \\
  24 - 35 & 6.6 & 0.91/0.42 & 144 - 155 & 8.5 & 0.32/0.64 & 274 - 284 & 7.6 & 0.56/0.38 \\
  35 - 41 & 7.6 & 0.77/0.35 & 155 - 165 & 8.5 & 0.38/0.49 & 284 - 294 & 8.5 & 0.53/0.35 \\
  41 - 48 & 6.6 & 0.65/0.34 & 165 - 175 & 7.6 & 0.66/0.42 & 294 - 305 & 8.5 & 0.49/0.40 \\
  48 - 54 & 7.6 & 0.51/0.33 & 175 - 184 & 8.5 & 0.79/0.60 & 305 - 311 & 9.5 & 0.47/0.53 \\
  54 - 65 & 8.5 & 0.37/0.34 & 184 - 194 & 7.6 & 0.86/0.76 & 311 - 318 & 12.3 & 0.48/0.52 \\
  65 - 75 & 9.5 & 0.36/0.39 & 194 - 204 & 8.5 & 0.86/0.70 & 318 - 324 & 11.4 & 0.52/0.62 \\
  75 - 85 & 12.3 & 0.41/0.46 & 204 - 215 & 12.3 & 0.77/0.65 & 324 - 335 & 9.5 & 0.54/0.80 \\
  85 - 94 & 17.1 & 0.44/0.57 & 215 - 222 & 12.3 & 0.65/0.72 & 335 - 345 & 8.5 & 0.65/0.94 \\
  94 - 104 & 12.3 & 0.59/0.63 & 222 - 228 & 13.3 & 0.59/0.67 & 345 - 355 & 7.6 & 0.78/0.85 \\
  104 - 114 & 8.5 & 0.85/0.73 & 228 - 234 & 11.4 & 0.58/0.58 & &&\\
  114 - 125 & 7.6 & 0.91/0.75 & 234 - 245 & 8.5 & 0.52/0.45 & &&\\
  \hline
\end{tabular}
\end{minipage}
\hfill
\caption{Illumination positions on the mirror during the spot scan measurement, the angular resolution (HPD) measured at 12 keV, and the corresponding spread of the slope error distribution at each position.}
\label{table:spotscan_hpd}
\end{table}
\subsection{Full-aperture X-ray Imaging Performance}\label{subsec:full}

\begin{figure*}[h!]
  \centering
  \includegraphics[width=0.9\linewidth]{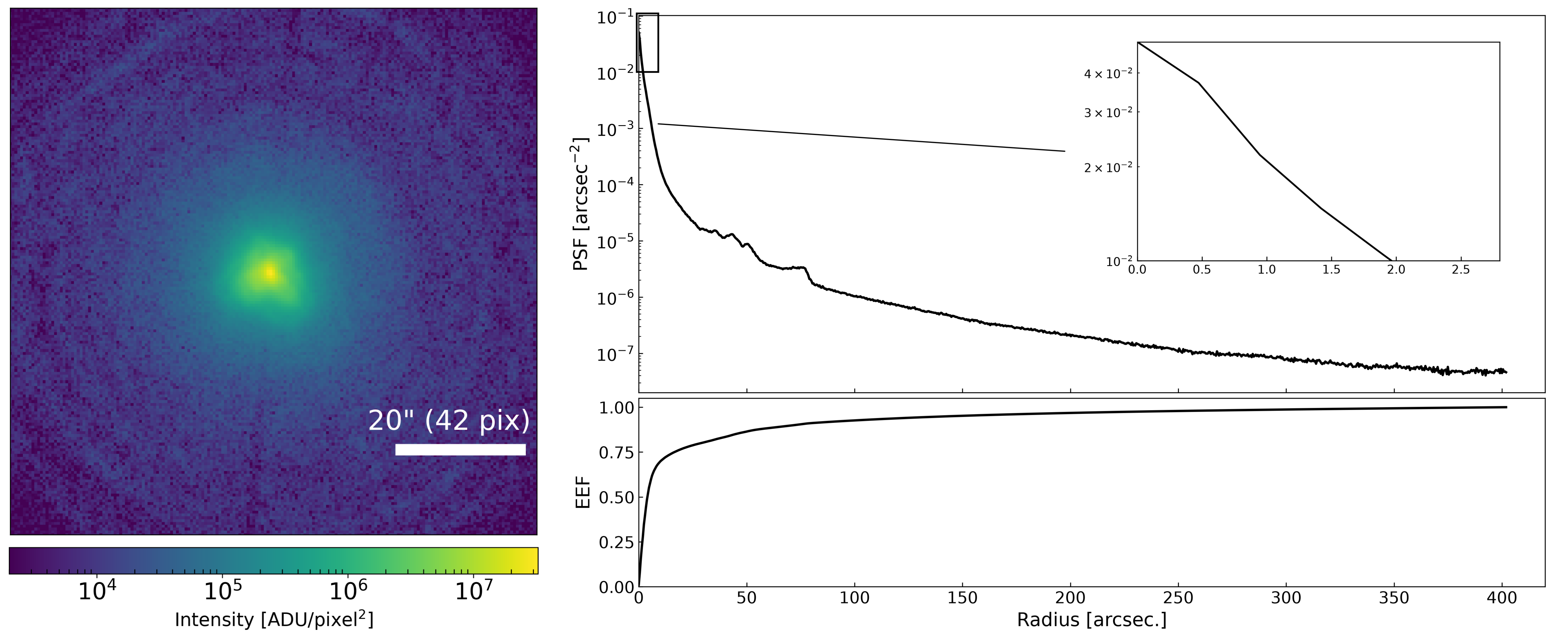}
  \caption{Result of the full-aperture illumination measurement at 12~keV using the HBX-KLAEES system.
  The focal-plane image is shown on the left, while the corresponding PSF and EEF are shown in the upper-right and lower-right panels, respectively.}
  \label{fig:figure_full}
\end{figure*}

The full-aperture imaging performance was evaluated at 12~keV using the HBX-KLAEES system. 
For background correction, a dark frame acquired with the upstream beam shutter closed was subtracted from the illuminated image.
Because a larger detector area was used for the full-aperture measurement, hot pixels were corrected by replacing each affected pixel with the mean value of the surrounding pixels. 
The FWHM and HPD were then evaluated from the PSF and EEF using the same procedure as described in Section~\ref{subsec:local}.
Figure~\ref{fig:figure_full} shows the focal-plane image obtained under full-aperture illumination, together with the corresponding PSF and EEF.
The measured full-aperture HPD was 8.5~arcsec, demonstrating sub-10-arcsec imaging performance when the entire reflecting aperture contributed
simultaneously to the focal image. 
This value is close to the HPDs of approximately 7--9 arcsec obtained over most of the mirror in the local spot scan measurements, demonstrating good consistency between the local and full-aperture measurements.
%
The full-aperture HPD improved from approximately 14~arcsec for the previous mirror reported by \citet{Fujii2026pasp} to 8.5~arcsec for the newly
fabricated mirror.

\section{Discussion} \label{sec:discussion}
\subsection{Origin of the Figure Errors in the Previous Mirror}\label{subsec:dis-previous}

The strong spatial correlation between the mandrel and mirror figure errors for the previous mandrel--mirror pair provides direct evidence that a substantial component of the mirror figure pattern originated from the mandrel. 
The Spearman rank correlation coefficient between the corresponding two-dimensional figure-error maps was \(r_s=0.80\). 
The correspondence is robust against intentional circumferential displacements of \(\pm1\) degree between the maps. 
This result quantitatively demonstrates that a substantial component of the spatial figure-error pattern of the previous mirror was inherited from the mandrel.
In contrast, the mandrel-to-mirror figure difference shows little correlation with the mandrel figure itself (\(r_s=0.11\)). 
Thus, although the mandrel figure was strongly reflected in the final mirror figure, the additional figure difference arising between the mandrel and final mirror showed little systematic correspondence with the local mandrel figure error. 
This distinction is important for understanding the previous mandrel--mirror pair. The results suggest that the final mirror figure consisted of at least two components: one retaining the spatial pattern of the mandrel and another arising between the mandrel and final mirror without a clear correspondence to the local mandrel figure-error pattern.
The latter component is not necessarily spatially uniform. 
As shown by the spatial correspondence of the branch-like populations in the correlation analysis, it includes localized figure deviations associated particularly with regions near the axial ends. 
Such localized deviations may contribute to the shoulder-like structures observed at several tens of arcseconds from the core in the measured PSFs. 
This interpretation is also consistent with simulation results obtained using our in-house ray-tracing simulator, as reported in \citet{Yoshida2024spie}, which showed that localized axial figure deviations can produce characteristic structures away from the PSF core.
\citet{Fujii2026pasp} traced the degradation of local X-ray performance to the axial figure errors of the final mirror; the present analysis extends this interpretation by identifying the mandrel as an important upstream origin of a substantial component of those errors.



\subsection{Change in the Relative Contributions to the Mirror Figure Error}\label{subsec:dis-new}

Improvement of the mandrel substantially reduced the figure errors of the newly fabricated mirror, as demonstrated by the reductions in the PV, RMS, and axial slope error widths described above.
More importantly, the spatial correlation analysis reveals a change in the relative contributions to the residual mirror figure error.
The mandrel--mirror correlation decreased from $r_{\rm s}=0.80$ for the previous mandrel--mirror pair to $r_{\rm s}=0.53$ for the new pair, while the mandrel remained essentially uncorrelated with the mandrel-to-mirror figure difference ($r_{\rm s}=0.11$ and $0.08$ for the previous and new mandrel--mirror pairs, respectively).

The lower mandrel--mirror correlation for the newly fabricated mirror does not necessarily indicate that the mandrel figure was less accurately replicated onto the mirror surface.
Because the amplitude and spatial variation of the mandrel figure errors were substantially reduced, residual figure-error components not directly associated with the mandrel surface became relatively more important in determining the spatial pattern of the final mirror figure.
Such components may include figure deviations introduced during the replication and subsequent fabrication processes, although minor contributions from measurement and spatial-registration uncertainties cannot be excluded.
These residual components can more readily alter the spatial rank ordering when the mandrel figure error itself is small, thereby reducing the Spearman rank correlation coefficient even though the absolute figure accuracy has improved.
A more direct indication of this change is provided by the relative magnitudes of the residual errors.
For the new mandrel--mirror pair, the axial slope error widths of the mandrel ($W68=1.6$ and $1.1$ arcsec for the paraboloidal and hyperboloidal surfaces, respectively) are comparable to those of the mandrel-to-mirror figure difference ($W68=1.7$ and $1.4$ arcsec, respectively).
Thus, after the substantial reduction of the mandrel figure errors, deviations arising between the mandrel and final mirror can no longer be regarded as negligible relative to the residual mandrel figure.
The branch-like populations associated with the axial ends, which are present in both the previous and new mirrors, further indicate that these residual deviations are not spatially uniform.
Their preferential occurrence near the axial ends may be related to the fabrication and separation processes.
For example, release of residual stress during separation from the mandrel may produce larger figure changes near the mirror edges, although the physical origin of these edge-related deviations remains to be investigated.
\subsection{Relation between Figure Errors and X-ray Imaging Performance}\label{subsec:dis-sld-hpd}

\begin{figure*}[h!]
  \centering
  \includegraphics[width=0.99\linewidth]{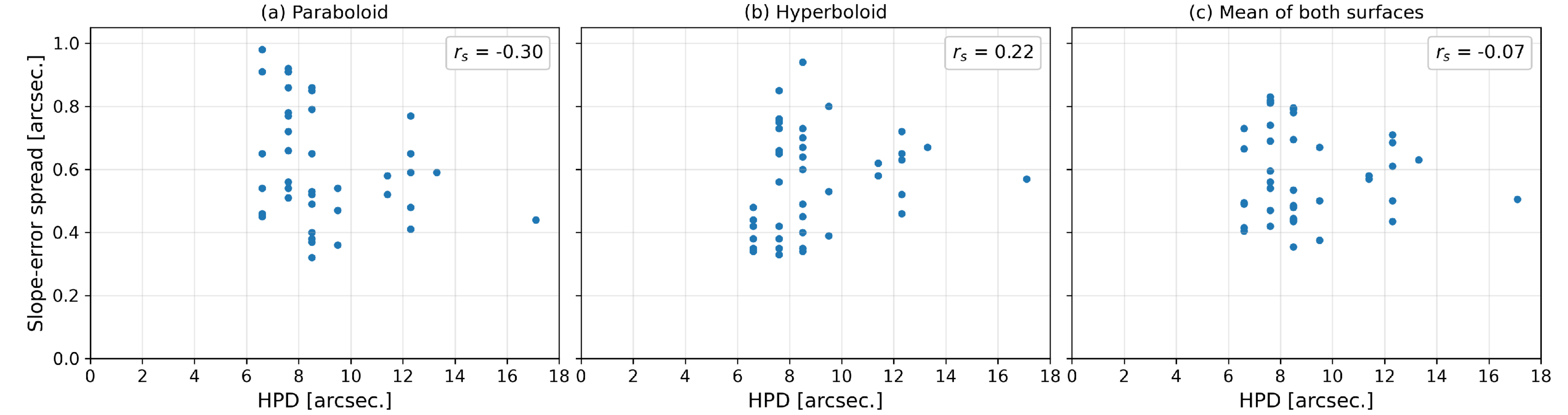}
  \caption{Relationships between the local HPD measured at 12 keV and the axial slope-error spread for the paraboloidal surface (left), hyperboloidal surface (middle), and their arithmetic mean (right). 
  The slope-error spread is defined as the median absolute deviation from the baseline, as in Table \ref{table:spotscan_hpd}. Spearman's rank correlation coefficients are indicated in each panel.}
  \label{fig:correlation_sle-hpd}
\end{figure*}

\begin{figure*}[h!]
  \centering
  \includegraphics[width=0.5\linewidth]{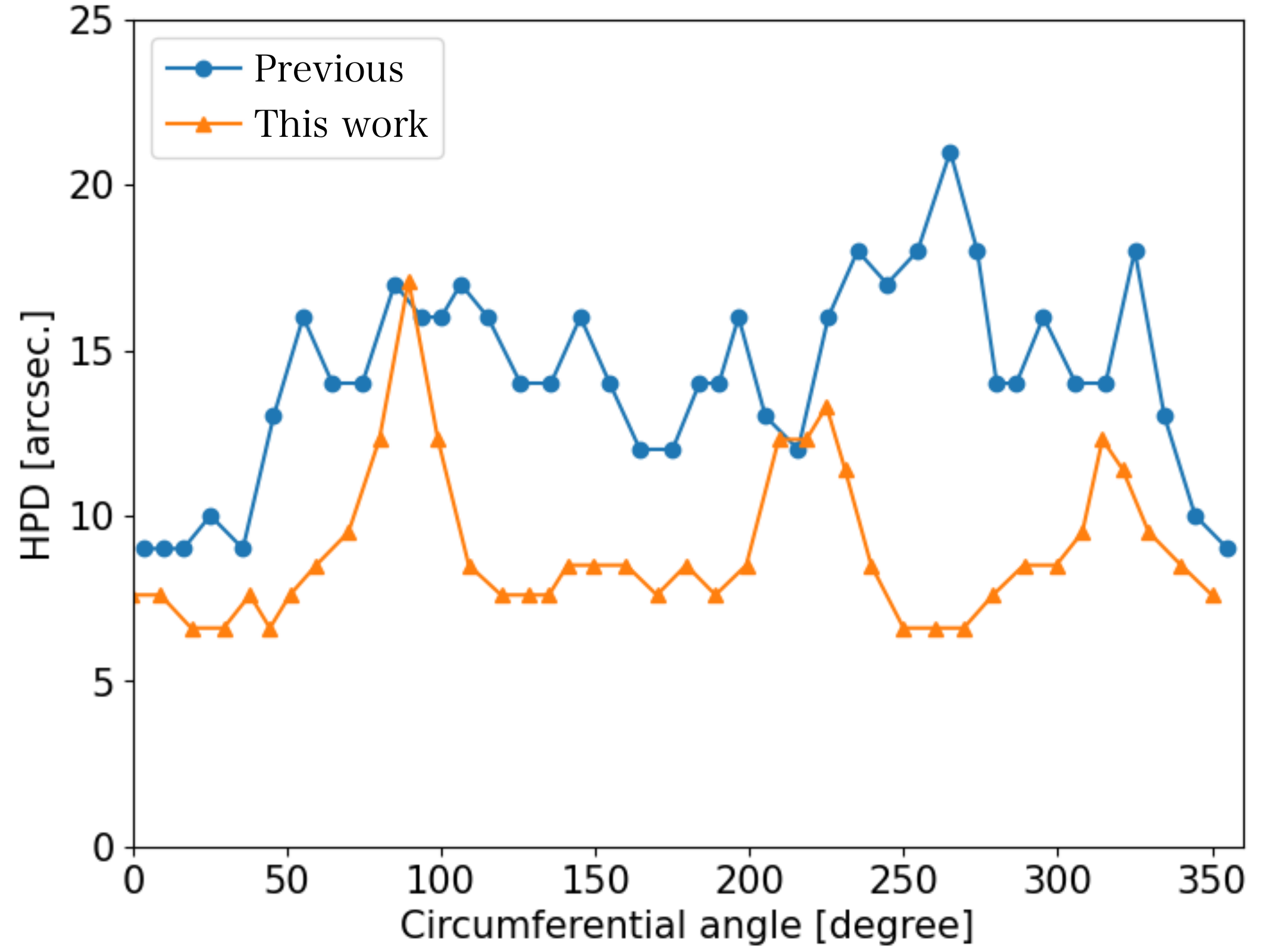}
  \caption{Azimuthal dependence of the local HPD measured at 12 keV for the previous (blue circles) and newly fabricated (orange triangles) mirrors.
  The azimuthal angle represents the circumferential position around the mirror. 
  The angle is measured clockwise from the horizontal rightward direction (0 deg) when viewed from the upstream (X-ray incident) side.
  The Spearman's rank correlation coefficient between the two HPD distributions is $r_s=\mathrm{0.22}$.}
  \label{fig:correlation_hpd-circumferential}
\end{figure*}

In the previous mirror, the local HPD showed a positive correlation with the axial slope error spread in the corresponding illuminated regions \citep{Fujii2026pasp}.
In contrast, such a relationship is no longer clearly apparent for the newly fabricated mirror.
Figure \ref{fig:correlation_sle-hpd} shows the relationship between the local HPD and the slope-error spread for the paraboloidal and hyperboloidal surfaces, as well as their arithmetic mean. Spearman's rank correlation coefficients were $r_s=-0.30$, $0.22$, and $-0.07$, respectively. Although a weak negative correlation coefficient was obtained for the paraboloidal surface, the mean slope-error spread showed little evidence of a correlation with the local HPD.
We also examined the diameter enclosing 75\% of the detected photons (75\%PD), which is more sensitive to the broader component of the PSF, but found no clear correlation with the axial slope error spread on either reflecting surface.
Figure \ref{fig:correlation_hpd-circumferential} compares the azimuthal dependence of the local HPD for the previous and newly fabricated mirrors. The newly fabricated mirror exhibits localized HPD enhancements near 90, 220, and 320 deg. 
Although similar azimuthal variations are present in the previous mirror, the correspondence between the two distributions is weak, with a Spearman's rank correlation coefficient of $r_s = 0.22$. Apart from the enhancement near 90 deg, the locations of the prominent HPD peaks differ between the two mirrors. 
There are no localized mandrel supports at these azimuthal positions, and the mirror release process is also considered unlikely to account for the observed HPD enhancements. 
The mandrel figure is refined through repeated cycles of figure measurements and corrective polishing, which can alter the spatial pattern of the residual figure errors. 
This interpretation is supported by the lack of a clear spatial correspondence between the figure-error patterns of the previous and improved mandrels, as shown in Figure \ref{fig:figure_2d-figureerror-map}.
These considerations suggest that the observed HPD enhancements are unlikely to arise from a common mechanical cause associated with fixed azimuthal positions.

\subsection{Application to the FOXSI-5 Flight Optics}\label{subsec:dis-foxsi5}

Based on the demonstrated X-ray imaging performance, the newly fabricated mirror was adopted for the soft X-ray optics of the fifth flight of the Focusing Optics X-ray Solar Imager (FOXSI-5) sounding rocket experiment \citep{foxsi5,Yoshida2026spie}.
The mirror was subsequently integrated into the MMA, followed by installation of the pre-collimator.
X-ray characterization after these integration steps showed no substantial degradation of the angular resolution compared with that measured for the mirror before integration.
No significant change in the imaging performance was observed after the subsequent vibration test.
These results demonstrate that the imaging performance achieved through the present precision-fabrication development was preserved through assembly, integration, and vibration testing of the flight MMA.
The completed MMA was subsequently flown aboard FOXSI-5, which was successfully launched in May 2026.

\section{Summary and Conclusions} \label{sec:summary-conclusion}

We fabricated and characterized a high-angular-resolution electroformed nickel X-ray mirror using a substantially improved precision mandrel. 
By directly comparing the corresponding two-dimensional figure-error distributions of the mandrels and replicated inner mirror surfaces for both the previous and newly fabricated mirrors, we extended our previous study \citep{Fujii2026pasp} to investigate the upstream origins of the mirror figure errors. 
The strong mandrel--mirror spatial correlation for the previous mirror indicates that a substantial component of its figure-error pattern was inherited from the mandrel. 
Following the improvement of the mandrel accuracy, the figure errors of the newly fabricated mirror were substantially reduced, while the mandrel--mirror spatial correlation became weaker. 
The residual figure deviations arising between the mandrel and final mirror became comparable in magnitude to the residual mandrel figure errors, suggesting that further improvement of the mirror figure will require better control of these deviations in addition to continued improvement of the mandrel accuracy.

The improved figure accuracy was accompanied by enhanced X-ray imaging performance, with a full-aperture HPD of 8.5 arcsec measured at 12 keV, compared with approximately 14 arcsec for the previous mirror. 
The newly fabricated mirror was subsequently integrated into the MMA for the FOXSI-5 soft X-ray telescope, and its imaging performance was maintained after installation of the pre-collimator and subsequent vibration testing. 
The completed MMA was successfully flown aboard FOXSI-5 in May 2026. 
These results demonstrate the applicability of the present precision electroforming approach to high-angular-resolution X-ray optics for space astronomy and highlight the importance of controlling both the mandrel figure and the deviations arising during subsequent mirror fabrication.

\begin{acknowledgments}
This work was supported by the Grants-in-Aid for Scientific Research (KAKENHI) from the Japan Society for the Promotion of Science (JSPS) under Grant Numbers JP22K18274, and JP20K20920 (IM), JP22H00134 and JP21KK0052 (NN), and JP23H00156 (HM).
Y.Y. and K.S. acknowledge support from the JSPS Research Fellowships for Young Scientists (DC1 and PD, respectively) under Grant Numbers JP26KJ1366 and JP26KJ0023, respectively.
Additional support was provided by JST SPRING (Grant Number JPMJSP2125) and the ISAS program for small-scale projects.
This work was also supported by the Iwadare Scholarship Foundation (MY).
KA acknowledges the generous support of the Hattori International Scholarship Foundation (HISF).
KA and KS also thank the “THERS Make New Standards Program for the Next Generation Researchers” for financial support.

\end{acknowledgments}
\bibliography{references_efoptics_mitsuishi_pasp}{}

@Inbook{Atkins2022book,
author="Atkins, Carolyn",
editor="Bambi, Cosimo
and Santangelo, Andrea",
title="Technologies for Advanced X-Ray Mirror Fabrication",
bookTitle="Handbook of X-ray and Gamma-ray Astrophysics",
year="2022",
publisher="Springer Nature Singapore",
address="Singapore",
pages="1--39",
isbn="978-981-16-4544-0",
doi="10.1007/978-981-16-4544-0_11-1",
url="https://doi.org/10.1007/978-981-16-4544-0_11-1"
}

@article{Singam2024spie,
author = {Panini Singam and Chet Speegle and Amy Meekham and Jeffery Kolodziejczak and David Banks and Wayne Baumgartner and Jessica Gaskin and Stephen Bongiorno and Brian Ramsey and David D. Smith and Danielle Gurgew and Nicholas Thomas},
title = {{Optimizing the electroforming process for full-shell X-ray optics}},
volume = {10},
journal = {Journal of Astronomical Telescopes, Instruments, and Systems},
number = {3},
publisher = {SPIE},
pages = {034001},
year = {2024},
doi = {10.1117/1.JATIS.10.3.034001},
URL = {https://doi.org/10.1117/1.JATIS.10.3.034001}
}

@article{Ramsey2022jatis,
author = {Brian D. Ramsey and Stephen D. Bongiorno and Jeffery K. Kolodziejczak and Kiranmayee Kilaru and Cheryl Alexander and Wayne H. Baumgartner and Shawn Breeding and Ronald F. Elsner and Shelley Le Roy and Jeff McCracken and Ikuyuki Mitsuishi and Stephen L. O'Dell and Steven D. Pavelitz and Jaganathan Ranganathan and Javier Sanchez and Chet O. Speegle and Nicholas Thomas and Bruce Weddendorf and Martin C. Weisskopf},
title = {{Optics for the imaging x-ray polarimetry explorer}},
volume = {8},
journal = {Journal of Astronomical Telescopes, Instruments, and Systems},
number = {2},
publisher = {SPIE},
pages = {024003},
year = {2022},
doi = {10.1117/1.JATIS.8.2.024003},
URL = {https://doi.org/10.1117/1.JATIS.8.2.024003}
}

@article{Predehl2021a&a,
	author = {{Predehl, P.} and {Andritschke, R.} and {Arefiev, V.} and {Babyshkin, V.} and {Batanov, O.} and {Becker, W.} and {Böhringer, H.} and {Bogomolov, A.} and {Boller, T.} and {Borm, K.} and {Bornemann, W.} and {Bräuninger, H.} and {Brüggen, M.} and {Brunner, H.} and {Brusa, M.} and {Bulbul, E.} and {Buntov, M.} and {Burwitz, V.} and {Burkert, W.} and {Clerc, N.} and {Churazov, E.} and {Coutinho, D.} and {Dauser, T.} and {Dennerl, K.} and {Doroshenko, V.} and {Eder, J.} and {Emberger, V.} and {Eraerds, T.} and {Finoguenov, A.} and {Freyberg, M.} and {Friedrich, P.} and {Friedrich, S.} and {Fürmetz, M.} and {Georgakakis, A.} and {Gilfanov, M.} and {Granato, S.} and {Grossberger, C.} and {Gueguen, A.} and {Gureev, P.} and {Haberl, F.} and {Hälker, O.} and {Hartner, G.} and {Hasinger, G.} and {Huber, H.} and {Ji, L.} and {Kienlin, A. v.} and {Kink, W.} and {Korotkov, F.} and {Kreykenbohm, I.} and {Lamer, G.} and {Lomakin, I.} and {Lapshov, I.} and {Liu, T.} and {Maitra, C.} and {Meidinger, N.} and {Menz, B.} and {Merloni, A.} and {Mernik, T.} and {Mican, B.} and {Mohr, J.} and {Müller, S.} and {Nandra, K.} and {Nazarov, V.} and {Pacaud, F.} and {Pavlinsky, M.} and {Perinati, E.} and {Pfeffermann, E.} and {Pietschner, D.} and {Ramos-Ceja, M. E.} and {Rau, A.} and {Reiffers, J.} and {Reiprich, T. H.} and {Robrade, J.} and {Salvato, M.} and {Sanders, J.} and {Santangelo, A.} and {Sasaki, M.} and {Scheuerle, H.} and {Schmid, C.} and {Schmitt, J.} and {Schwope, A.} and {Shirshakov, A.} and {Steinmetz, M.} and {Stewart, I.} and {Strüder, L.} and {Sunyaev, R.} and {Tenzer, C.} and {Tiedemann, L.} and {Trümper, J.} and {Voron, V.} and {Weber, P.} and {Wilms, J.} and {Yaroshenko, V.}},
	title = {The eROSITA X-ray telescope on SRG},
	DOI= "10.1051/0004-6361/202039313",
	url= "https://doi.org/10.1051/0004-6361/202039313",
	journal = {A\&A},
	year = 2021,
	volume = 647,
	pages = "A1",
}

@inproceedings{Ramsey1999spie,
author = {Brian D. Ramsey and Darell E. Engelhaupt and Chet O. Speegle and Stephen L. O'Dell and Robert A. Austin and Jeffery J. Kolodziejczak and Martin C. Weisskopf},
title = {{HERO program: high-energy replicated optics for a hard-x-ray balloon payload}},
volume = {3765},
booktitle = {EUV, X-Ray, and Gamma-Ray Instrumentation for Astronomy X},
editor = {Oswald H. W. Siegmund and Kathryn A. Flanagan},
organization = {International Society for Optics and Photonics},
publisher = {SPIE},
pages = {816 -- 821},
year = {1999},
doi = {10.1117/12.366569},
URL = {https://doi.org/10.1117/12.366569}
}

@inproceedings{Pareschi2005spie,
author = {G. Pareschi and S. Basso and O. Citterio and M. Ghigo and F. Mazzoleni and D. Spiga and W. Burkert and M. Freyberg and G. D. Hartner and G. Conti and E. Mattaini and G. Grisoni and G. Valsecchi and B. Negri and G. Parodi and A. Marzorati and P. dell'Acqua},
title = {{Development of grazing-incidence multilayer mirrors by direct Ni electroforming replication: a status report}},
volume = {5900},
booktitle = {Optics for EUV, X-Ray, and Gamma-Ray Astronomy II},
editor = {Oberto Citterio and Stephen L. O'Dell},
organization = {International Society for Optics and Photonics},
publisher = {SPIE},
pages = {590008},
year = {2005},
doi = {10.1117/12.617362},
URL = {https://doi.org/10.1117/12.617362}
}

@article{Gubarev2010nims,
title = {An investigation of differential deposition for figure corrections in full-shell grazing-incidence X-ray optics},
journal = {Nuclear Instruments and Methods in Physics Research Section A: Accelerators, Spectrometers, Detectors and Associated Equipment},
volume = {616},
number = {2},
pages = {273-276},
year = {2010},
note = {X-Ray Mirror},
issn = {0168-9002},
doi = {https://doi.org/10.1016/j.nima.2009.10.161},
url = {https://www.sciencedirect.com/science/article/pii/S0168900209021135},
author = {M.V. Gubarev and K. Kilaru and B.D. Ramsey}
}

@INPROCEEDINGS{Aschenbach2002spie,
       author = {{Aschenbach}, Bernd},
        title = "{In-orbit performance of the XMM-Newton x-ray telescopes: images and spectra}",
    booktitle = {X-Ray Optics for Astronomy: Telescopes, Multilayers, Spectrometers, and Missions},
         year = 2002,
       editor = {{Gorenstein}, Paul and {Hoover}, Richard B.},
       series = {Society of Photo-Optical Instrumentation Engineers (SPIE) Conference Series},
       volume = {4496},
        month = jan,
        pages = {8-22},
          doi = {10.1117/12.454367},
archivePrefix = {arXiv},
       eprint = {astro-ph/0109375},
 primaryClass = {astro-ph},
       adsurl = {https://ui.adsabs.harvard.edu/abs/2002SPIE.4496....8A}
}

@ARTICLE{Kilaru2019jatis,
       author = {{Kilaru}, Kiranmayee and {Ramsey}, Brian D. and {Baumgartner}, Wayne H. and {Bongiorno}, Stephen D. and {Broadway}, David M. and {Champey}, Patrick R. and {Davis}, Jacqueline M. and {O'Dell}, Stephen L. and {Elsner}, Ronald F. and {Gaskin}, Jessica A. and {Johnson}, Samantha and {Kolodziejczak}, Jeffery K. and {Roberts}, Oliver J. and {Swartz}, Douglas A. and {Weisskopf}, Martin C.},
        title = "{Full-shell x-ray optics development at NASA Marshall Space Flight Center}",
      journal = {Journal of Astronomical Telescopes, Instruments, and Systems},
         year = 2019,
        month = apr,
       volume = {5},
          eid = {021010},
        pages = {021010},
          doi = {10.1117/1.JATIS.5.2.021010},
       adsurl = {https://ui.adsabs.harvard.edu/abs/2019JATIS...5b1010K}
}

@ARTICLE{Mimura2018rsi,
       author = {{Mimura}, Hidekazu and {Takei}, Yoshinori and {Kume}, Takehiro and {Takeo}, Yoko and {Motoyama}, Hiroto and {Egawa}, Satoru and {Matsuzawa}, Yusuke and {Yamaguchi}, Gota and {Senba}, Yasunori and {Kishimoto}, Hikaru and {Ohashi}, Haruhiko},
        title = "{Fabrication of a precise ellipsoidal mirror for soft X-ray nanofocusing}",
      journal = {Review of Scientific Instruments},
         year = 2018,
        month = sep,
       volume = {89},
       number = {9},
          eid = {093104},
        pages = {093104},
          doi = {10.1063/1.5035323},
       adsurl = {https://ui.adsabs.harvard.edu/abs/2018RScI...89i3104M}
}

@ARTICLE{Kume2019rsi,
       author = {{Kume}, Takehiro and {Takei}, Yoshinori and {Egawa}, Satoru and {Motoyama}, Hiroto and {Takeo}, Yoko and {Yamaguchi}, Gota and {Mimura}, Hidekazu},
        title = "{Development of electroforming process for soft x-ray ellipsoidal mirror}",
      journal = {Review of Scientific Instruments},
         year = 2019,
        month = feb,
       volume = {90},
       number = {2},
          eid = {021718},
        pages = {021718},
          doi = {10.1063/1.5057432},
       adsurl = {https://ui.adsabs.harvard.edu/abs/2019RScI...90b1718K}
}

@ARTICLE{Takeo2020apphl,
       author = {{Takeo}, Yoko and {Suzuki}, Akihiro and {Motoyama}, Hiroto and {Takei}, Yoshinori and {Kume}, Takehiro and {Matsuzawa}, Yusuke and {Senba}, Yasunori and {Kishimoto}, Hikaru and {Ohashi}, Haruhiko and {Mimura}, Hidekazu},
        title = "{Soft x-ray nanobeam formed by an ellipsoidal mirror}",
      journal = {Applied Physics Letters},
         year = 2020,
        month = mar,
       volume = {116},
       number = {12},
          eid = {121102},
        pages = {121102},
          doi = {10.1063/1.5144932},
       adsurl = {https://ui.adsabs.harvard.edu/abs/2020ApPhL.116l1102T}
}

@ARTICLE{Takeo2020apphl2,
       author = {{Takeo}, Yoko and {Motoyama}, Hiroto and {Shimamura}, Takenori and {Kimura}, Takashi and {Kume}, Takehiro and {Matsuzawa}, Yusuke and {Saito}, Takahiro and {Imamura}, Yoichi and {Miyashita}, Hiroaki and {Hiraguri}, Kentaro and {Hashizume}, Hirokazu and {Senba}, Yasunori and {Kishimoto}, Hikaru and {Ohashi}, Haruhiko and {Mimura}, Hidekazu},
        title = "{A highly efficient nanofocusing system for soft x rays}",
      journal = {Applied Physics Letters},
         year = 2020,
        month = oct,
       volume = {117},
       number = {15},
          eid = {151104},
        pages = {151104},
          doi = {10.1063/5.0027118},
       adsurl = {https://ui.adsabs.harvard.edu/abs/2020ApPhL.117o1104T}
}

@INPROCEEDINGS{Yamaguchi2020spie,
       author = {{Yamaguchi}, Gota and {Kume}, Takehiro and {Takeo}, Yoko and {Matsuzawa}, Yusuke and {Hiraguri}, Kentaro and {Imamura}, Yoichi and {Miyashita}, Hiroaki and {Tamura}, Keisuke and {Takehara}, Yusuke and {Takigawa}, Ayumu and {Kanoh}, Tetsuo and {Tachibana}, Kenji and {Mitsuishi}, Ikuyuki and {Hashizume}, Hirokazu and {Mimura}, Hidekazu},
        title = "{Development of precise electroformed Wolter mirror for X-ray astronomy}",
    booktitle = {Advances in X-Ray/EUV Optics and Components XV},
         year = 2020,
       series = {Society of Photo-Optical Instrumentation Engineers (SPIE) Conference Series},
       volume = {11491},
        month = sep,
          eid = {114910K},
        pages = {114910K},
          doi = {10.1117/12.2569834},
       adsurl = {https://ui.adsabs.harvard.edu/abs/2020SPIE11491E..0KY}
}

@inproceedings{Ito2021spieproc,
author = {Akinari Ito and Gota Yamaguchi and Takafumi Watanabe and Yoko Takeo and Takehiro Kume and Yusuke Matsuzawa and Takahiro Saito and Kentaro Hiraguri and Keisuke Tamura and Ayumu Takigawa and Tetsuo Kanoh and Naoki Ishida and Ikuyuki Mitsuishi and Hirokazu Hashizume and Hidekazu Mimura},
title = {{Electrodeposition simulation for fabricating Wolter mirrors of x-ray telescopes}},
volume = {11837},
booktitle = {Advances in X-Ray/EUV Optics and Components XVI},
editor = {Christian Morawe and Ali M. Khounsary and Shunji Goto},
organization = {International Society for Optics and Photonics},
publisher = {SPIE},
pages = {118370D},
year = {2021},
doi = {10.1117/12.2595124},
URL = {https://doi.org/10.1117/12.2595124}
}

@article{Yamaguchi2023rsi,
    author = {Yamaguchi, Gota and Matsuzawa, Yusuke and Kume, Takehiro and Imamura, Yoichi and Miyashita, Hiroaki and Ito, Akinari and Sakuta, Koki and Ampuku, Kazuki and Fujii, Ryuto and Hiraguri, Kentaro and Hashizume, Hirokazu and Mitsuishi, Ikuyuki and Mimura, Hidekazu},
    title = {Efficient and precise fabrication of Wolter type-I x-ray mirrors via nickel electroforming replication using quartz glass mandrels},
    journal = {Review of Scientific Instruments},
    volume = {94},
    number = {12},
    pages = {124501},
    year = {2023},
    month = {12},
    issn = {0034-6748},
    doi = {10.1063/5.0160262},
    url = {https://doi.org/10.1063/5.0160262},
    eprint = {https://pubs.aip.org/aip/rsi/article-pdf/doi/10.1063/5.0160262/18260147/124501_1_5.0160262.pdf},
}

@INPROCEEDINGS{Krucker2013spie,
       author = {{Krucker}, S{\"a}m. and {Christe}, Steven and {Glesener}, Lindsay and {Ishikawa}, Shinnosuke and {Ramsey}, Brian and {Gubarev}, Mikhail and {Saito}, Shinya and {Takahashi}, Tadayuki and {Watanabe}, Shin and {Tajima}, Hiroyasu and {Tanaka}, Takaaki and {Turin}, Paul and {Glaser}, David and {Fermin}, Jose and {Lin}, Robert P.},
        title = "{The focusing optics x-ray solar imager (FOXSI): instrument and first flight}",
    booktitle = {Solar Physics and Space Weather Instrumentation V},
         year = 2013,
       editor = {{Fineschi}, Silvano and {Fennelly}, Judy},
       series = {Society of Photo-Optical Instrumentation Engineers (SPIE) Conference Series},
       volume = {8862},
        month = sep,
          eid = {88620R},
        pages = {88620R},
          doi = {10.1117/12.2024277},
       adsurl = {https://ui.adsabs.harvard.edu/abs/2013SPIE.8862E..0RK}
}

@inproceedings{Yoshida2024spie,
author = {Yusuke Yoshida and Koki Sakuta and Kazuki Ampuku and Ryuto Fujii and Kumiko Okada and Keitoku Yoshihira and Tetsuo Kano and Naoki Ishida and Wataru Kato and Yoshitaka Inoue and Keisuke Tamura and Kikuko Miyata and Noriyuki Narukage and Gota Yamaguchi and Shunsuke Ito and Shutaro Mohri and Takehiro Kume and Yusuke Matsuzawa and Yoichi Imamura and Takahiro Saito and Kentaro Hiraguri and Hirokazu Hashizume and Hidekazu Mimura and Ikuyuki Mitsuishi},
title = {{Development of x-ray optics for the solar flare sounding rocket FOXSI-4: ray-tracing simulation}},
volume = {13093},
booktitle = {Space Telescopes and Instrumentation 2024: Ultraviolet to Gamma Ray},
editor = {Jan-Willem A. den Herder and Shouleh Nikzad and Kazuhiro Nakazawa},
organization = {International Society for Optics and Photonics},
publisher = {SPIE},
pages = {130936W},
year = {2024},
doi = {10.1117/12.3018175},
URL = {https://doi.org/10.1117/12.3018175}
}

@inproceedings{Ampuku2024spie,
author = {Kazuki Ampuku and Koki Sakuta and Ryuto Fujii and Yusuke Yoshida and Kumiko Okada and Keitoku Yoshihira and Tetsuo Kano and Naoki Ishida and Wataru Kato and Yoshitaka Inoue and Keisuke Tamura and Kikuko Miyata and Noriyuki Narukage and Gota Yamaguchi and Shunsuke Ito and Shutaro Mohri and Takehiro Kume and Yusuke Matsuzawa and Yoichi Imamura and Takahiro Saito and Kentaro Hiraguri and Hirokazu Hashizume and Hidekazu Mimura and Ikuyuki Mitsuishi},
title = {{Development of x-ray optics for the solar flare sounding rocket FOXSI-4: ground calibration}},
volume = {13093},
booktitle = {Space Telescopes and Instrumentation 2024: Ultraviolet to Gamma Ray},
editor = {Jan-Willem A. den Herder and Shouleh Nikzad and Kazuhiro Nakazawa},
organization = {International Society for Optics and Photonics},
publisher = {SPIE},
pages = {1309323},
year = {2024},
doi = {10.1117/12.3018109},
URL = {https://doi.org/10.1117/12.3018109}
}

@article{astropy:2013,
Adsurl = {http://adsabs.harvard.edu/abs/2013A%26A...558A..33A},
Archiveprefix = {arXiv},
Author = {{Astropy Collaboration} and {Robitaille}, T.~P. and {Tollerud}, E.~J. and {Greenfield}, P. and {Droettboom}, M. and {Bray}, E. and {Aldcroft}, T. and {Davis}, M. and {Ginsburg}, A. and {Price-Whelan}, A.~M. and {Kerzendorf}, W.~E. and {Conley}, A. and {Crighton}, N. and {Barbary}, K. and {Muna}, D. and {Ferguson}, H. and {Grollier}, F. and {Parikh}, M.~M. and {Nair}, P.~H. and {Unther}, H.~M. and {Deil}, C. and {Woillez}, J. and {Conseil}, S. and {Kramer}, R. and {Turner}, J.~E.~H. and {Singer}, L. and {Fox}, R. and {Weaver}, B.~A. and {Zabalza}, V. and {Edwards}, Z.~I. and {Azalee Bostroem}, K. and {Burke}, D.~J. and {Casey}, A.~R. and {Crawford}, S.~M. and {Dencheva}, N. and {Ely}, J. and {Jenness}, T. and {Labrie}, K. and {Lim}, P.~L. and {Pierfederici}, F. and {Pontzen}, A. and {Ptak}, A. and {Refsdal}, B. and {Servillat}, M. and {Streicher}, O.},
Doi = {10.1051/0004-6361/201322068},
Eid = {A33},
Eprint = {1307.6212},
Journal = {\aap},
Month = oct,
Pages = {A33},
Primaryclass = {astro-ph.IM},
Title = {{Astropy: A community Python package for astronomy}},
Volume = 558,
Year = 2013}

@ARTICLE{astropy:2018,
       author = {{Astropy Collaboration} and {Price-Whelan}, A.~M. and
         {Sip{\H{o}}cz}, B.~M. and {G{\"u}nther}, H.~M. and {Lim}, P.~L. and
         {Crawford}, S.~M. and {Conseil}, S. and {Shupe}, D.~L. and
         {Craig}, M.~W. and {Dencheva}, N. and {Ginsburg}, A. and {Vand
        erPlas}, J.~T. and {Bradley}, L.~D. and {P{\'e}rez-Su{\'a}rez}, D. and
         {de Val-Borro}, M. and {Aldcroft}, T.~L. and {Cruz}, K.~L. and
         {Robitaille}, T.~P. and {Tollerud}, E.~J. and {Ardelean}, C. and
         {Babej}, T. and {Bach}, Y.~P. and {Bachetti}, M. and {Bakanov}, A.~V. and
         {Bamford}, S.~P. and {Barentsen}, G. and {Barmby}, P. and
         {Baumbach}, A. and {Berry}, K.~L. and {Biscani}, F. and {Boquien}, M. and
         {Bostroem}, K.~A. and {Bouma}, L.~G. and {Brammer}, G.~B. and
         {Bray}, E.~M. and {Breytenbach}, H. and {Buddelmeijer}, H. and
         {Burke}, D.~J. and {Calderone}, G. and {Cano Rodr{\'\i}guez}, J.~L. and
         {Cara}, M. and {Cardoso}, J.~V.~M. and {Cheedella}, S. and {Copin}, Y. and
         {Corrales}, L. and {Crichton}, D. and {D'Avella}, D. and {Deil}, C. and
         {Depagne}, {\'E}. and {Dietrich}, J.~P. and {Donath}, A. and
         {Droettboom}, M. and {Earl}, N. and {Erben}, T. and {Fabbro}, S. and
         {Ferreira}, L.~A. and {Finethy}, T. and {Fox}, R.~T. and
         {Garrison}, L.~H. and {Gibbons}, S.~L.~J. and {Goldstein}, D.~A. and
         {Gommers}, R. and {Greco}, J.~P. and {Greenfield}, P. and
         {Groener}, A.~M. and {Grollier}, F. and {Hagen}, A. and {Hirst}, P. and
         {Homeier}, D. and {Horton}, A.~J. and {Hosseinzadeh}, G. and {Hu}, L. and
         {Hunkeler}, J.~S. and {Ivezi{\'c}}, {\v{Z}}. and {Jain}, A. and
         {Jenness}, T. and {Kanarek}, G. and {Kendrew}, S. and {Kern}, N.~S. and
         {Kerzendorf}, W.~E. and {Khvalko}, A. and {King}, J. and {Kirkby}, D. and
         {Kulkarni}, A.~M. and {Kumar}, A. and {Lee}, A. and {Lenz}, D. and
         {Littlefair}, S.~P. and {Ma}, Z. and {Macleod}, D.~M. and
         {Mastropietro}, M. and {McCully}, C. and {Montagnac}, S. and
         {Morris}, B.~M. and {Mueller}, M. and {Mumford}, S.~J. and {Muna}, D. and
         {Murphy}, N.~A. and {Nelson}, S. and {Nguyen}, G.~H. and
         {Ninan}, J.~P. and {N{\"o}the}, M. and {Ogaz}, S. and {Oh}, S. and
         {Parejko}, J.~K. and {Parley}, N. and {Pascual}, S. and {Patil}, R. and
         {Patil}, A.~A. and {Plunkett}, A.~L. and {Prochaska}, J.~X. and
         {Rastogi}, T. and {Reddy Janga}, V. and {Sabater}, J. and
         {Sakurikar}, P. and {Seifert}, M. and {Sherbert}, L.~E. and
         {Sherwood-Taylor}, H. and {Shih}, A.~Y. and {Sick}, J. and
         {Silbiger}, M.~T. and {Singanamalla}, S. and {Singer}, L.~P. and
         {Sladen}, P.~H. and {Sooley}, K.~A. and {Sornarajah}, S. and
         {Streicher}, O. and {Teuben}, P. and {Thomas}, S.~W. and
         {Tremblay}, G.~R. and {Turner}, J.~E.~H. and {Terr{\'o}n}, V. and
         {van Kerkwijk}, M.~H. and {de la Vega}, A. and {Watkins}, L.~L. and
         {Weaver}, B.~A. and {Whitmore}, J.~B. and {Woillez}, J. and
         {Zabalza}, V. and {Astropy Contributors}},
        title = "{The Astropy Project: Building an Open-science Project and Status of the v2.0 Core Package}",
      journal = {\aj},
         year = 2018,
        month = sep,
       volume = {156},
       number = {3},
          eid = {123},
        pages = {123},
          doi = {10.3847/1538-3881/aabc4f},
archivePrefix = {arXiv},
       eprint = {1801.02634},
 primaryClass = {astro-ph.IM},
       adsurl = {https://ui.adsabs.harvard.edu/abs/2018AJ....156..123A}
}

@ARTICLE{astropy:2022,
       author = {{Astropy Collaboration} and {Price-Whelan}, Adrian M. and {Lim}, Pey Lian and {Earl}, Nicholas and {Starkman}, Nathaniel and {Bradley}, Larry and {Shupe}, David L. and {Patil}, Aarya A. and {Corrales}, Lia and {Brasseur}, C.~E. and {N{"o}the}, Maximilian and {Donath}, Axel and {Tollerud}, Erik and {Morris}, Brett M. and {Ginsburg}, Adam and {Vaher}, Eero and {Weaver}, Benjamin A. and {Tocknell}, James and {Jamieson}, William and {van Kerkwijk}, Marten H. and {Robitaille}, Thomas P. and {Merry}, Bruce and {Bachetti}, Matteo and {G{"u}nther}, H. Moritz and {Aldcroft}, Thomas L. and {Alvarado-Montes}, Jaime A. and {Archibald}, Anne M. and {B{'o}di}, Attila and {Bapat}, Shreyas and {Barentsen}, Geert and {Baz{'a}n}, Juanjo and {Biswas}, Manish and {Boquien}, M{'e}d{'e}ric and {Burke}, D.~J. and {Cara}, Daria and {Cara}, Mihai and {Conroy}, Kyle E. and {Conseil}, Simon and {Craig}, Matthew W. and {Cross}, Robert M. and {Cruz}, Kelle L. and {D'Eugenio}, Francesco and {Dencheva}, Nadia and {Devillepoix}, Hadrien A.~R. and {Dietrich}, J{"o}rg P. and {Eigenbrot}, Arthur Davis and {Erben}, Thomas and {Ferreira}, Leonardo and {Foreman-Mackey}, Daniel and {Fox}, Ryan and {Freij}, Nabil and {Garg}, Suyog and {Geda}, Robel and {Glattly}, Lauren and {Gondhalekar}, Yash and {Gordon}, Karl D. and {Grant}, David and {Greenfield}, Perry and {Groener}, Austen M. and {Guest}, Steve and {Gurovich}, Sebastian and {Handberg}, Rasmus and {Hart}, Akeem and {Hatfield-Dodds}, Zac and {Homeier}, Derek and {Hosseinzadeh}, Griffin and {Jenness}, Tim and {Jones}, Craig K. and {Joseph}, Prajwel and {Kalmbach}, J. Bryce and {Karamehmetoglu}, Emir and {Ka{l}uszy{'n}ski}, Miko{l}aj and {Kelley}, Michael S.~P. and {Kern}, Nicholas and {Kerzendorf}, Wolfgang E. and {Koch}, Eric W. and {Kulumani}, Shankar and {Lee}, Antony and {Ly}, Chun and {Ma}, Zhiyuan and {MacBride}, Conor and {Maljaars}, Jakob M. and {Muna}, Demitri and {Murphy}, N.~A. and {Norman}, Henrik and {O'Steen}, Richard and {Oman}, Kyle A. and {Pacifici}, Camilla and {Pascual}, Sergio and {Pascual-Granado}, J. and {Patil}, Rohit R. and {Perren}, Gabriel I. and {Pickering}, Timothy E. and {Rastogi}, Tanuj and {Roulston}, Benjamin R. and {Ryan}, Daniel F. and {Rykoff}, Eli S. and {Sabater}, Jose and {Sakurikar}, Parikshit and {Salgado}, Jes{'u}s and {Sanghi}, Aniket and {Saunders}, Nicholas and {Savchenko}, Volodymyr and {Schwardt}, Ludwig and {Seifert-Eckert}, Michael and {Shih}, Albert Y. and {Jain}, Anany Shrey and {Shukla}, Gyanendra and {Sick}, Jonathan and {Simpson}, Chris and {Singanamalla}, Sudheesh and {Singer}, Leo P. and {Singhal}, Jaladh and {Sinha}, Manodeep and {Sip{H{o}}cz}, Brigitta M. and {Spitler}, Lee R. and {Stansby}, David and {Streicher}, Ole and {{{S}}umak}, Jani and {Swinbank}, John D. and {Taranu}, Dan S. and {Tewary}, Nikita and {Tremblay}, Grant R. and {Val-Borro}, Miguel de and {Van Kooten}, Samuel J. and {Vasovi{'c}}, Zlatan and {Verma}, Shresth and {de Miranda Cardoso}, Jos{'e} Vin{'i}cius and {Williams}, Peter K.~G. and {Wilson}, Tom J. and {Winkel}, Benjamin and {Wood-Vasey}, W.~M. and {Xue}, Rui and {Yoachim}, Peter and {Zhang}, Chen and {Zonca}, Andrea and {Astropy Project Contributors}},
        title = "{The Astropy Project: Sustaining and Growing a Community-oriented Open-source Project and the Latest Major Release (v5.0) of the Core Package}",
      journal = {\apj},
         year = 2022,
        month = aug,
       volume = {935},
       number = {2},
          eid = {167},
        pages = {167},
          doi = {10.3847/1538-4357/ac7c74},
archivePrefix = {arXiv},
       eprint = {2206.14220},
 primaryClass = {astro-ph.IM},
       adsurl = {https://ui.adsabs.harvard.edu/abs/2022ApJ...935..167A}
}

@Article{Hunter:2007,
  Author    = {Hunter, J. D.},
  Title     = {Matplotlib: A 2D graphics environment},
  Journal   = {Computing in Science \& Engineering},
  Volume    = {9},
  Number    = {3},
  Pages     = {90--95},
  publisher = {IEEE COMPUTER SOC},
  doi       = {10.1109/MCSE.2007.55},
  year      = 2007
}

@Article{harris2020array,
 title         = {Array programming with {NumPy}},
 author        = {Charles R. Harris and K. Jarrod Millman and St{\'{e}}fan J.
                 van der Walt and Ralf Gommers and Pauli Virtanen and David
                 Cournapeau and Eric Wieser and Julian Taylor and Sebastian
                 Berg and Nathaniel J. Smith and Robert Kern and Matti Picus
                 and Stephan Hoyer and Marten H. van Kerkwijk and Matthew
                 Brett and Allan Haldane and Jaime Fern{\'{a}}ndez del
                 R{\'{i}}o and Mark Wiebe and Pearu Peterson and Pierre
                 G{\'{e}}rard-Marchant and Kevin Sheppard and Tyler Reddy and
                 Warren Weckesser and Hameer Abbasi and Christoph Gohlke and
                 Travis E. Oliphant},
 year          = {2020},
 month         = sep,
 journal       = {Nature},
 volume        = {585},
 number        = {7825},
 pages         = {357--362},
 doi           = {10.1038/s41586-020-2649-2},
 publisher     = {Springer Science and Business Media {LLC}},
 url           = {https://doi.org/10.1038/s41586-020-2649-2}
}

@software{reback2020pandas,
    author       = {The pandas development team},
    title        = {pandas-dev/pandas: Pandas},
    month        = feb,
    year         = 2020,
    publisher    = {Zenodo},
    version      = {latest},
    doi          = {10.5281/zenodo.3509134},
    url          = {https://doi.org/10.5281/zenodo.3509134}
}

@ARTICLE{2020SciPy-NMeth,
  author  = {Virtanen, Pauli and Gommers, Ralf and Oliphant, Travis E. and
            Haberland, Matt and Reddy, Tyler and Cournapeau, David and
            Burovski, Evgeni and Peterson, Pearu and Weckesser, Warren and
            Bright, Jonathan and {van der Walt}, St{\'e}fan J. and
            Brett, Matthew and Wilson, Joshua and Millman, K. Jarrod and
            Mayorov, Nikolay and Nelson, Andrew R. J. and Jones, Eric and
            Kern, Robert and Larson, Eric and Carey, C J and
            Polat, {\.I}lhan and Feng, Yu and Moore, Eric W. and
            {VanderPlas}, Jake and Laxalde, Denis and Perktold, Josef and
            Cimrman, Robert and Henriksen, Ian and Quintero, E. A. and
            Harris, Charles R. and Archibald, Anne M. and
            Ribeiro, Ant{\^o}nio H. and Pedregosa, Fabian and
            {van Mulbregt}, Paul and {SciPy 1.0 Contributors}},
  title   = {{{SciPy} 1.0: Fundamental Algorithms for Scientific
            Computing in Python}},
  journal = {Nature Methods},
  year    = {2020},
  volume  = {17},
  pages   = {261--272},
  adsurl  = {https://rdcu.be/b08Wh},
  doi     = {10.1038/s41592-019-0686-2},
}

@article{Fujii2026pasp,
doi = {10.1088/1538-3873/ae3b74},
url = {https://doi.org/10.1088/1538-3873/ae3b74},
year = {2026},
month = {apr},
publisher = {The Astronomical Society of the Pacific},
volume = {138},
number = {4},
pages = {045003},
author = {Fujii, Ryuto and Sakuta, Koki and Ampuku, Kazuki and Yoshida, Yusuke and Yoshihara, Makoto and Takigawa, Ayumu and Yoshihira, Keitoku and Kano, Tetsuo and Ishida, Naoki and Narukage, Noriyuki and Tamura, Keisuke and Miyata, Kikuko and Yamaguchi, Gota and Takano, Hidekazu and Kohmura, Yoshiki and Mohri, Shutaro and Kume, Takehiro and Matsuzawa, Yusuke and Imamura, Yoichi and Saito, Takahiro and Hiraguri, Kentaro and Hashizume, Hirokazu and Mimura, Hidekazu and Mitsuishi, Ikuyuki},
title = {Development of Electroformed X-Ray Optics Bridging Synchrotron Radiation Technology and Space Astronomy},
journal = {Publications of the Astronomical Society of the Pacific}
}

@inproceedings{Buitrago-Casas2021spie,
author = {Juan Camilo Buitrago-Casas and Juliana Vievering and Sophie Musset and Lindsay Glesener and P.S. Athiray and Wayne Baumgartner and Stephen Bongiorno and Patrick Champey and Steven Christe and Sasha Courtade and Gregory Dalton and Jessie Duncan and Kelsey Gilchrist and Shin-nosuke Ishikawa and Christine Jhabvala and Hunter Kanniainen and S{¥"a}m Krucker and Kyle Gregory and Juan Carlos Martinez Oliveros and Jeff McCracken and Ikuyuki Mitsuishi and Noriyuki Narukage and Athanasios Pantazides and Eliad Peretz and Savannah Perez-Piel and Aruna Ramanayaka and Brian Ramsey and Danny Ryan and Sabrina Savage and Tadayuki Takahashi and Shin Watanabe and Amy Winebarger and Yixian Zhang},
title = {{FOXSI-4: the high resolution focusing X-ray rocket payload to observe a solar flare}},
volume = {11821},
booktitle = {UV, X-Ray, and Gamma-Ray Space Instrumentation for Astronomy XXII},
editor = {Oswald H. Siegmund},
organization = {International Society for Optics and Photonics},
publisher = {SPIE},
pages = {118210L},
year = {2021},
doi = {10.1117/12.2594701},
URL = {https://doi.org/10.1117/12.2594701}
}

@inproceedings{Yoshida2026spie,
author = {Yusuke Yoshida and Kazuki Ampuku and Koki Sakuta and Makoto Yoshihara and Keitoku Yoshihira and Ryoma Tanaka and Tetsuo Kano and Naoki Ishida and Wataru Kato and Yoshitaka Inoue and Keisuke Tamura and Takashi Okajima and Kikuko Miyata and Noriyuki Narukage and Gota Yamaguchi and Shutaro Mohri and Satoru Egawa and Hiroto Motoyama and Takehiro Kume and Yusuke Matsuzawa and Yoichi Imamura and Takahiro Saito and Kentaro Hiraguri and Hirokazu Hashizume and Hidekazu Mimura and Ikuyuki Mitsuishi},
title = {{Development of x-ray optics for the solar flare sounding rocket FOXSI-5}},
volume = {14146},
booktitle = {Space Telescopes and Instrumentation 2026: Ultraviolet to Gamma Ray},
editor = {Shouleh Nikzad and Kazuhiro Nakazawa and Marco Feroci},
organization = {International Society for Optics and Photonics},
publisher = {SPIE},
pages = {141464K},
year = {2026},
doi = {10.1117/12.3100626},
URL = {https://doi.org/10.1117/12.3100626}
}

@MISC{foxsi5,
       author = {{Buitrago Casas}, Juan Camilo},
        title = "{FOXSI-5: A Rocket Re-flight for High Resolution Focused X-ray Observations of the Sun}",
 howpublished = {NASA Proposal ID. 23-HLCAS23-0009},
         year = 2023,
        month = jan,
        pages = {9},
       adsurl = {https://ui.adsabs.harvard.edu/abs/2023hlca.prop....9B}
}
\bibliographystyle{aasjournalv7}



\end{document}